\documentclass[a4paper,x11names,12pt]{article}
\normalsize
\pdfoutput=1

\usepackage[utf8]{inputenc}
\usepackage{lmodern}
\usepackage[T1]{fontenc} 
\usepackage{microtype} 

\usepackage[a4paper, left=25mm, right=25mm, top=30mm, bottom=25mm]{geometry} 
\usepackage{fancyhdr} 
\usepackage{abstract}

\usepackage{titlesec} 
\titleformat{\section}[block]{\large\bfseries\centering}{\thesection}{1em}{} 
\titleformat{\subsection}[block]{\bfseries}{\thesubsection}{1em}{} 

\usepackage{xcolor}

\usepackage{cite}
\usepackage{hyperref}
\hypersetup{
      colorlinks=true,
      linkcolor=Blue4,
      citecolor=Red4,
      urlcolor=Green4,
      linktoc=page
}

\usepackage{enumerate}
\usepackage{graphicx}
\usepackage[percent]{overpic}
\usepackage{tikz}
\usepackage{subcaption}
\usepackage[many]{tcolorbox}
\usepackage{booktabs}
\tcbset{highlight math style={enhanced,
  colframe=DeepSkyBlue4,colback=Honeydew1}}

\usepackage{amsmath,amssymb,slashed,mathbbol,mathtools}
\usepackage{amsthm}
\numberwithin{equation}{section}
\usepackage{slashed}

\DeclareSymbolFontAlphabet{\amsmathbb}{AMSb}%

\newcommand{\beq}{\begin{equation}}
\newcommand{\eeq}{\end{equation}}
\newcommand{\be}{\begin{equation}}
\newcommand{\ee}{\end{equation}}
\newcommand{\bea}{\begin{eqnarray}}
\newcommand{\eea}{\end{eqnarray}}
\newcommand{\beas}{\begin{eqnarray*}}
\newcommand{\eeas}{\end{eqnarray*}}

\newcommand{\bquo}{\begin{quote}}
\newcommand{\enqu}{\end{quote}}
\renewcommand{\(}{\begin{equation}}
\renewcommand{\)}{\end{equation}}

\def\e{\mathrm{e}}

\newcommand{\Z}{\ensuremath{\mathbb Z}}

\newcommand{\dd}{\mathrm{d}}
\newcommand{\f}[2]{\frac{#1}{#2}}

\title{\fontsize{19pt}{23pt}\selectfont\textbf{
Late-Time Saturation of Black Hole Complexity}\vspace{5mm}}

\author{
\normalsize{
\href{mailto:ffg1m24@soton.ac.uk}{\bf Fri{\dh}rik Freyr Gautason$^{1,2}$}, \href{mailto:vym1@hi.is}{\bf Vyshnav Mohan$^{1,3}$}, \textbf{and} \href{mailto:lth@hi.is}{\bf L\'arus Thorlacius$^{1,3}$}}\\[5mm]
{\normalsize $^1$ Science Institute, University of Iceland}\\[-1mm]
{\normalsize Dunhaga 3, 107 Reykjav{\'i}k, Iceland}\\[5mm]
{\normalsize $^2\,$STAG Research centre \& Mathematical Sciences, University of Southampton,}\\[-1mm]
{\normalsize Highfield, Southampton SO17 1BJ, U.K.}\\[5mm]
{\normalsize $^3\,$Institutionen för fysik och astronomi, Uppsala Universitet,}\\[-1mm]
{\normalsize Box 803, SE-751 08 Uppsala, Sweden}\\[2mm]
}

\date{}

  {\itshape}
  {}

\begin{document} 
\begin{flushright}
\footnotesize UUITP-06/25\\
\normalsize
\end{flushright}
\vspace{1cm}
{\let\newpage\relax\hypersetup{urlcolor=black}\maketitle}
\thispagestyle{empty}


\vspace{0.6cm}
\begin{abstract}
\noindent 
The holographic complexity of a static spherically symmetric black hole, defined as the volume of an extremal surface, grows linearly with time at late times in general relativity. The growth comes from a region at a constant transverse area inside the black hole and continues forever in the classical theory. In this region the volume complexity of any spherically symmetric black hole in $d+1$ spacetime dimensions reduces to a geodesic length in an effective two-dimensional JT-gravity theory. The length in JT-gravity has been argued to saturate at very late times via non-perturbative corrections obtained from a random matrix description of the gravity theory. The same argument, applied to our effective JT-gravity description of the volume complexity, leads to complexity saturation at times of exponential order in the Bekenstein-Hawking entropy of a $d+1$-dimensional black hole. Along the way, we explore a simple toy model for complexity growth, based on a discretisation of Nielsen complexity geometry, that can be analytically shown to exhibit the expected late-time complexity saturation.
\end{abstract}
\newpage
\tableofcontents

\newpage

\section{Introduction}
\label{sec:intro}
In this paper, we consider the very late time behaviour of the holographic complexity of black holes in $d+1\geq 2$ spacetime dimensions. Our starting point is the well-known Complexity=Volume (CV) prescription \cite{Susskind:2014rva,Stanford:2014jda} for a static spherically symmetric black hole, where the complexity is defined to be proportional to the extremal volume of spacelike slices that pass through the black hole interior and are anchored at fixed area spherical surfaces outside the black hole. 

According to the principle of black hole complementarity  \cite{Susskind:1993if}, a black hole appears to distant observers as a quantum system with a finite number of degrees of freedom that account for the Bekenstein-Hawking entropy of the black hole. The associated quantum dynamics, which must be sufficiently chaotic to scramble quantum information on a relatively short timescale, then provides a holographic dual representation of the spacetime geometry and matter inside the black hole. In particular, the holographic complexity is a geometric feature of the bulk gravity that corresponds to quantum computational complexity in a quantum circuit description of the black hole dynamics (see \cite{Susskind:2018pmk} for a review). 

The quantum computational complexity of a chaotic system exhibits universal features that are independent of the exact details of the system. In particular, complexity grows linearly for an extended time for any macroscopic system, following an initial period of scrambling. The growth rate is proportional to the number of degrees of freedom of the system. Moreover, complexity saturates at very late times of the order of exponential of the entropy \cite{Susskind:2015toa,Brown:2016wib, Balasubramanian:2019wgd,Susskind:2020wwe,Balasubramanian:2021mxo,Haferkamp:2021uxo}. This ramp-plateau structure is expected to be a universal behaviour of all finite-dimensional chaotic systems including black holes.
The volume of extremal surfaces, however, grows forever and never saturates and it is not {\it a priori\/} clear that a semiclassical prescription can reproduce the very late time plateau. We will explore this question and show that complexity, defined through the CV prescription, saturates at the expected time scales in any static, spherically symmetric black hole in any number of dimensions.

Our main argument for the saturation of holographic black hole complexity follows from the following observations. Susskind and Stanford \cite{Stanford:2014jda} noted that the volume of extremal surfaces can be expressed as a geodesic length in an auxiliary two-dimensional metric. Additionally, at late times, the growth of the extremal volume comes almost exclusively from a region inside the black hole where the transverse area remains constant. We refer to the limiting constant area slice at infinite time as the \textit{accumulation surface}. In the spacetime region near the accumulation surface, the auxiliary two-dimensional metric is negatively curved, and since the accumulation surface sits at a constant radial coordinate, it will have a fixed negative curvature in this region. Consequently, we can rephrase the late-time growth of the extremal volume as the growth of geodesic length in a Jackiw-Teitelboim (JT) gravity theory \cite{Jackiw:1984je,Teitelboim:1983ux} with an AdS$_2$ length scale derived from the higher-dimensional theory. 

The length of geodesics in classical JT gravity was considered in \cite{Brown:2018bms} and found to grow linearly for perpetuity at the expected rate. Subsequently, a quantum representation of geodesic length was defined and studied in \cite{Iliesiu:2021ari,Alishahiha:2022kzc,Iliesiu:2024cnh,Stanford:2022fdt}  and it was shown that the {\it quantum\/} length in fact saturates at very late times. 
We argue that one can use the quantum description of geodesic length in JT-gravity to arrive at a non-perturbative description of the complexity of a $d+1$-dimensional black hole at late times. This is because quantum corrections to the classical volume are exponentially suppressed by the area of the transverse $(d-1)$-sphere in Planck units. Since the transverse area is macroscopic for the entire extremal volume surface, it would seem that quantum corrections do not play a role at all. However, the strong suppression is eventually compensated by the growing span of the Einstein-Rosen bridge and quantum corrections cannot be neglected at sufficiently late times. The crucial point is that the dominant contribution to the volume increase, that comes from the region near the accumulation surface, is captured by the emergent JT description. On the other hand, the contribution to the extremal volume from regions away from the accumulation surface does not grow with time and furthermore the transverse area is even larger in those regions, leading to even stronger suppression of quantum corrections.
Translating the JT-gravity results back to higher dimensions, we find that complexity saturates at times of the order $e^{O(S)}$, where $S$ is the entropy of the higher-dimensional black hole (refer to table~\ref{table1} below for explicit expressions). Such a time scale is expected from general arguments \cite{Barbon:2019wsy}. 

The rest of the paper is organized as follows. 
To provide a simple, intuitive picture of the saturation of quantum computational complexity, we begin in Section~\ref{randomwalksec} by considering a toy model that can be viewed as a discretisation of Nielsen complexity geometry \cite{Nielsen:2005mkt} as a high-dimensional hypercube. The evolution of the quantum state is modelled by a random walk on the hypercube, where the walker starts from an arbitrarily chosen origin and can advance to a nearest neighbor at each time step. The average graph distance of the walker from the origin will then exhibit the expected behaviour of complexity. After a long period of linear growth, the distance saturates as the walker reaches the ``bulk'' of the hypercube, where the total distance from the origin is equally likely to increase or decrease with the next step.
Section~\ref{CVsection} briefly reviews the Complexity=Volume prescription. In section~\ref{2dsection}, we reduce the volume calculation to a geodesic length calculation in an effective two-dimensional metric, following \cite{Stanford:2014jda} and in section~\ref{JTgravitysec}, we show how the emergent two-dimensional metric on the relevant part of the extremal slice reduces to that of JT-gravity at late times. In section~\ref{saturationfromJTsection}, we use JT-gravity results to obtain the time scales at which the complexity saturates. We close with a brief discussion in Section~\ref{discussionsection} and 
some explicit examples are worked out in Appendix~\ref{flatspacesection}.

\section{A toy model for complexity}
\label{randomwalksec}
Before embarking on explorations in holographic complexity, which composes the bulk of this paper, we discuss some general features of quantum computational complexity by studying a simplified toy model. Recall that circuit complexity of an operator is a measure of how many `simple' operations are required to build that target operator from a given reference operator. As such, complexity depends on the precise definition of the `simple' operations (usually called gate set), the reference, and the tolerance by which we allow the constructed operator to deviate from the target under consideration. Our interest will be in the growth of complexity with time under Hamiltonian evolution of some initial state and in particular its late time behaviour. It turns out, that different detailed definitions of complexity lead to broadly speaking the same behaviour on long time scales. In particular, complexity is expected to grow linearly for a long time, at a rate proportional to the number of degrees of freedom that participate in the quantum dynamics, and then eventually saturate. 

Relative complexity of two operators defines a distance on the set of all unitary operators acting on the Hilbert space, which inspired Nielsen to define complexity geometry \cite{Nielsen:2005mkt,Nielsen:2006cea,Dowling:2006tnk} which involves a choice of a metric on the group manifold of all unitary transformations. Complexity is then defined to be the shortest geodesic distance between two points on the manifold as measured by the complexity metric. Once again, different choices of metric will assign different values to the complexity but a broad class of metrics is expected to lead to the generic behaviour of the complexity discussed above.

In \cite{Lin:2018cbk} Lin introduced a simplified notion of complexity which involves choosing a large but finite subgroup $G$ of the unitary group that has a generating set $S$. Such a finite group has a natural graph structure called a Cayley graph, where each element $g\in G$ is assigned a vertex and for every $g\in G$ and $s\in S$ there is a (directed) edge between the vertices $g$ and $gs$. If the generating set $S$ is closed under inversion the graph is undirected. Relative complexity between two group elements is the shortest graph distance between them, {\it i.e.} the minimal number of edges that connect the two corresponding vertices in the Cayley graph. Replacing the continuous Nielsen geometry by a discrete Cayley graph can be visualized as picking a particularly large tolerance $\epsilon$ where we are not so concerned with constructing precisely the target operator but are content with reaching its approximate neighbourhood (as measured by the inner product).

\subsection{Random walk on a hypercube}
A particularly simple choice of a finite group is $\Z_2^D$ with the generating set 
\be
S=\{(1,0,0,\cdots),(0,1,0,\cdots),\cdots \}\,.
\ee
The resulting Cayley graph is a high-dimension hypercube where each corner of the cube labels an operator and each edge connected to a given corner represents one of $D$ simple operations in this model. We will model the time evolution by a random walk on the hypercube starting from some reference vertex. We are interested in studying the graph distance between the current location of the walker and its initial position as a proxy for the complexity growth. 
\begin{figure}[h!]
\centering
\includegraphics[width=0.4\textwidth]{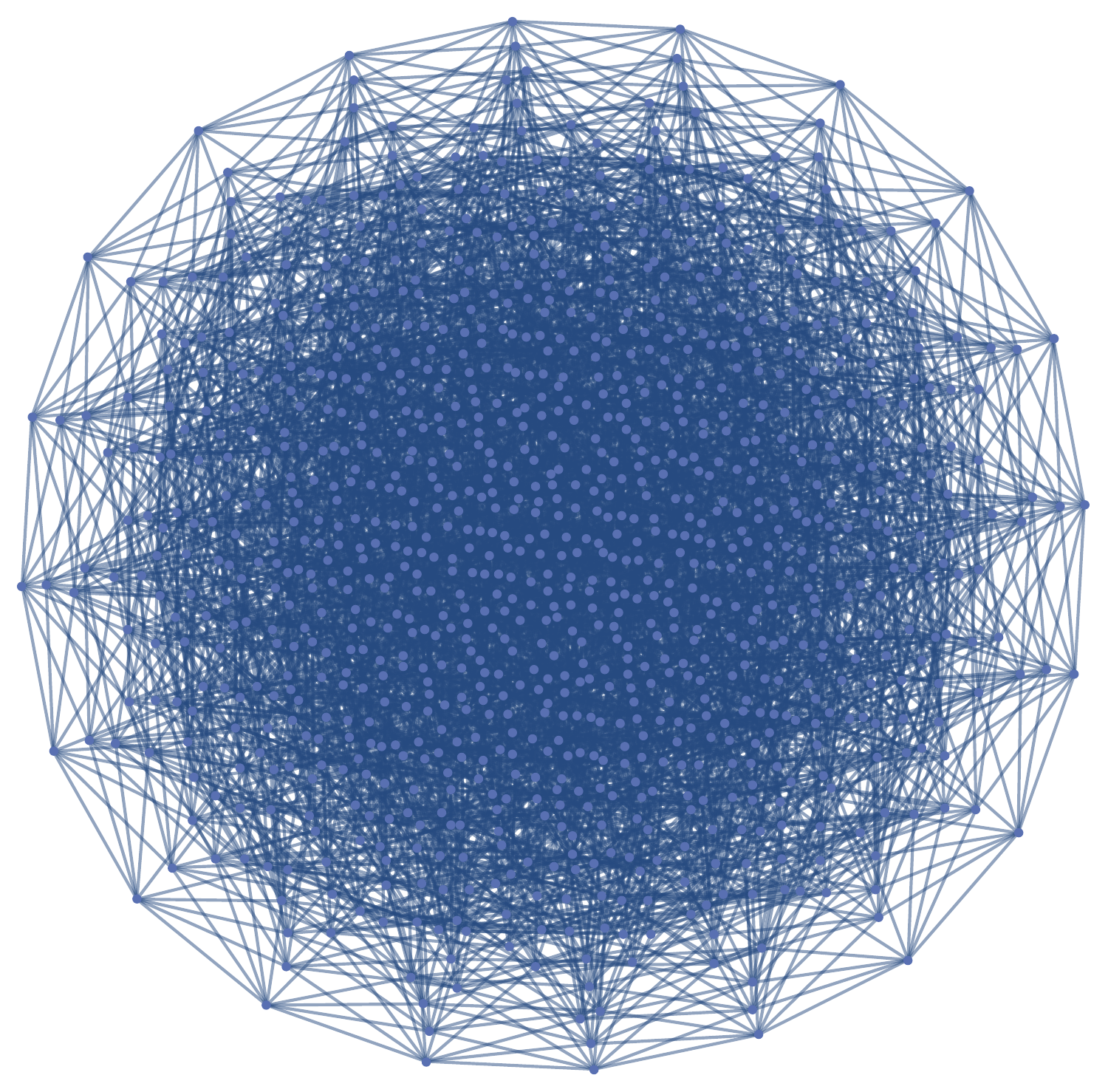}
\caption{\label{10Dcube}\small Graph representation of a ten-dimensional cube.}
\end{figure}
This is a classic problem that has been studied for instance in \cite{https://doi.org/10.1002/rsa.3240010105}, which we follow. The position of the walker can be parametrized by a $D$-dimensional vector where each entry takes the values $0$ or $1$. For simplicity, we take the reference state and the start point of the random walk to be at the origin ${\bf x} = {\bf 0}$. The random walk we consider is such that the walker takes one step to one of its nearest neighbours or stays put with equal probability $P = 1/(D+1)$. This means that at each time step the position vector ${\bf x}$ either stays unchanged or one of its entries is switched from 0 to 1 or vice versa. The graph distance is given by $|{\bf x}| = x_1 + x_2 +x_3 +\cdots$.

As explained in \cite{https://doi.org/10.1002/rsa.3240010105}, in the continuum version of the random walk, each entry in the position vector independently satisfies
\be\label{probofeachentry}
P[x_i =1] = \frac12 \Big( 1-\e^{-2\tau/D} \Big)\,,
\ee
where $\tau$ is the time variable that counts the time steps. This can be understood as follows; for large $D$ the variable $x_i$ can be thought of as counting the times that entry has been changed modulo 2. So it counts the occurrences of the random jump to that entry. Since each occurrence is rare and is independent of the previous ones it follows a Poisson distribution with the expectation $\lambda = \tau/(D+1)\approx \tau/D$. The probability of $k$ occurrences is 
$\lambda^k\e^{-\lambda}/k!$ and the probability for our variable $x_i$ to change an odd number of times is given by the sum
\be
P[x_i =1] = \sum_{k=0}^\infty\frac{\lambda^{2k+1}}{(2k+1)!}\e^{-\lambda} = \e^{-\lambda}\sinh \lambda\,,
\ee
which agrees with \eqref{probofeachentry}.

The result \eqref{probofeachentry} shows that on long time scales the uniform distribution is reached $P[x_i=1]=1/2$. That means we are equally likely to find the random walker on any vertex of the cube. In other words the walker has completely forgotten its initial location. For the uniform distribution it is simple to compute the expected distance to the origin and all higher moments. These are defined as
\be
\langle x^d\rangle = \frac{1}{2^D}\sum_{x=1}^D \binom{D}{x} x^d\,,
\ee
where we use the graph distance (here denoted by simply $x$) to perform the expectation value. The moments can be computed using the exponential generating function,
\be
G_D(y)\equiv \sum_{d=0}^\infty \langle x^d\rangle \frac{y^d}{d!}= \frac{1}{2^D}\sum_{x=1}^D \binom{D}{x} (\e^{y})^x = \Big(\frac{1+\e^{y}}{2}\Big)^D\,,
\ee
by taking $y$-derivatives and setting $y$ to zero at the end. In this way we compute
\be
\langle 1\rangle =1\,,\quad \langle x \rangle = \frac{D}{2}\,,\quad \langle x^2\rangle = \frac{D(D+1)}{4}\,, \quad \langle x^3\rangle = \frac{D^2(D+3)}{8}\,,\quad \cdots
\ee
and the standard deviation is $\Delta \equiv \sqrt{\langle x^2\rangle-\langle x\rangle^2}= \sqrt{D}/2$.

Let us now return to the time-dependent problem. Using the result \eqref{probofeachentry}, we can compute the expected distance as a function of time
\be\label{expdist}
C(\tau) \equiv \langle x(\tau)\rangle = \frac{D}{2} \Big( 1-\e^{-2\tau/D} \Big)\,,
\ee
which saturates at late times to the expected distance $D/2$. This saturation can be estimated to happen when $\langle x(\tau)\rangle$ is one standard deviation from the mean, {\it i.e.} when
\be
\langle x(\tau_\text{sat})\rangle = \frac{D}{2}-\frac{\sqrt{D}}{2}\,, 
\ee
or
\be
\tau_\text{sat} = \frac14 D \log D\,.
\ee
After the saturation time is reached, the random walker explores the cube according to the uniform distribution. Randomly it will come back to its original starting position with probability $2^{-D}$. The expected time between visiting the same point twice is therefore $\tau_P\sim2^{D}$. This serves as the Ponc\'are recurrence time in our model.

\begin{figure}[h!]
\centering
\includegraphics[width=0.8\textwidth]{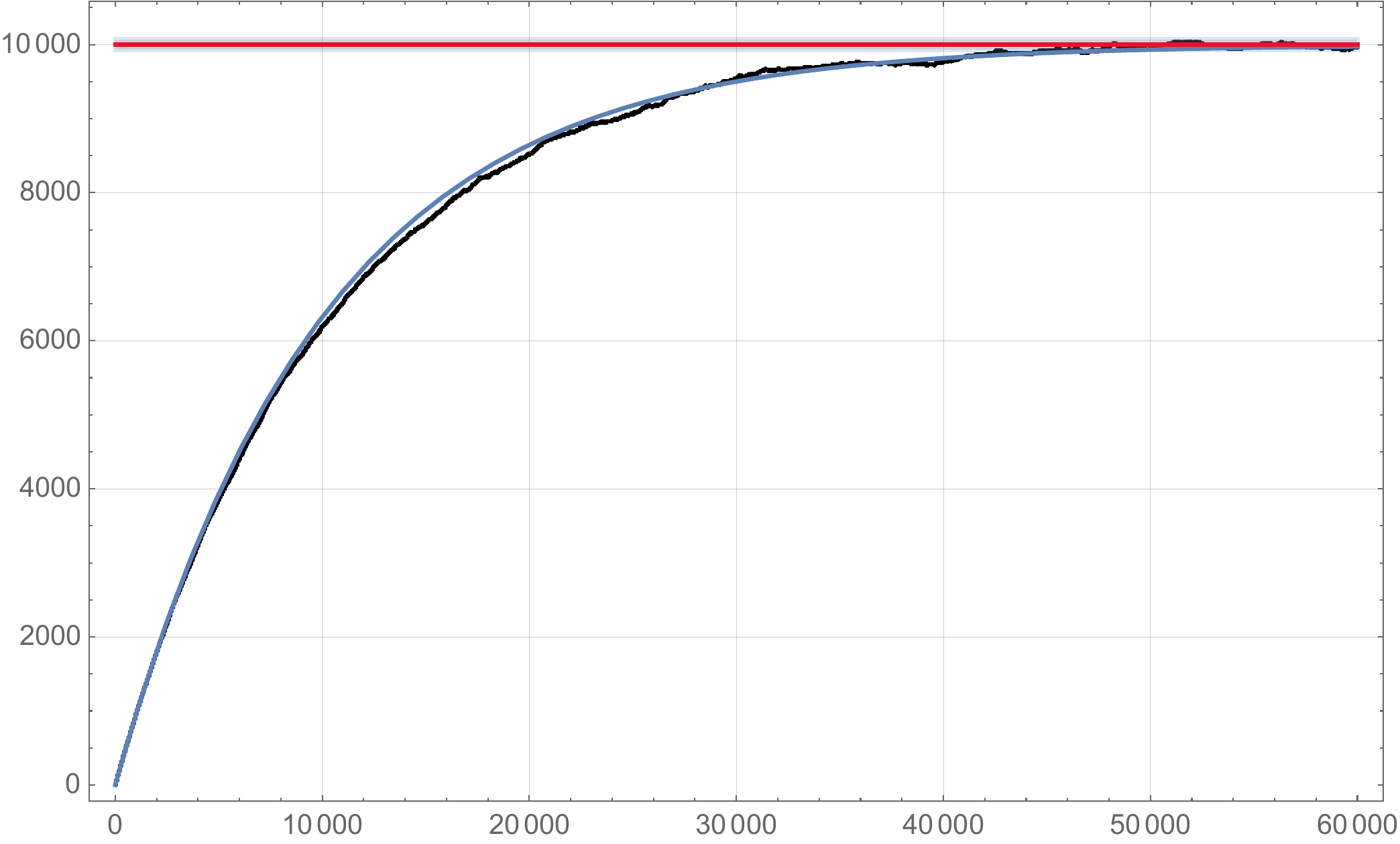}
\caption{\label{SimulationPlot}\small The black curve shows the distance of a simulated random walker on a $D=20000$ dimensional cube from its starting point -- the origin. In blue we show the expected distance \eqref{expdist}. We also display the saturation value at $D/2$ and shade in the standard deviation.}
\end{figure}

In order to make a connection to holographic complexity, we will take the dimensionality of our hypercube to be of exponential order in the black hole entropy $D=\e^{\alpha\,S}$ where $\alpha$ is a constant. Since it is mainly the exponential dependence on $S$ that is important, we set $\alpha=1$ to keep the notation simple.  Notice that the function $C(\tau)$ in \eqref{expdist} grows linearly at early times,
\be
C(\tau) =\tau + {\cal O}(\tau^2)\,,
\ee
just like the volume of the Einstein-Rosen bridge does. However, as discussed at the end of Section~\ref{CVsection}, the growth rate of the volume complexity of a black hole is proportional to the black hole entropy $S$ in dimensionless Rindler-like time units. Indeed the choice of the time coordinate has been somewhat arbitrary so far, where each time step represents a step of the random walk. To match with black hole complexity we choose $t = \tau/S$ as our dimensionless time coordinate.\footnote{For notational simplicity, we have set another dimensionless constant to one in our definition of the Rindler time.} Then
\be
C(t) \equiv \frac{\e^S}{2} \Big( 1-\e^{-2t S \e^{-S}} \Big)\,.
\ee
The two important time-scales in these variables are the saturation time and Poincar\'e recurrence time, given by
\be
t_\text{sat} = \frac14 \e^{S}\,,\qquad t_P \sim S \,2^{\e^S}\,.
\ee
In Figure~\ref{SimulationPlot} we display a sample simulation of a random walk and show that it follows the expected distance \eqref{expdist} quite closely.

Note that the random walker has reached approximately half the maximal graph distance from the origin on the hypercube at the saturation time. From then on the graph distance is equally likely to increase or decrease in a given step. We expect similar behaviour in a more realistic description of quantum complexity, {\it i.e.} that saturation occurs well before the complexity reaches its maximal possible value and that complexity equilibrium can be explored by studying a uniform 
distribution on the Hilbert space. 
Decreasing complexity has been linked to opaqueness of a black hole horizon \cite{Susskind:2015toa} and in our simple
model such a complexity `firewall'  occurs with probability one-half after the complexity has saturated. This matches recent results in JT gravity \cite{Blommaert:2024ftn,Iliesiu:2024cnh}, where the probability of a firewall is also found to be one-half at late times.

\section{Growth of Extremal Volume Surfaces}
\label{CVsection}
Let us briefly review the Complexity=Volume proposal \cite{Susskind:2014rva, Stanford:2014jda}. Consider a static, spherically symmetric, black hole solution in $d+1$ dimensions, whose metric is given by
\bea
d s^2=-f(\rho) \dd{t}^2+\frac{\dd \rho^2}{f(\rho)}+\rho^2 \dd{}\Omega_{d-1}^2\,,\label{staticbhmetric}
\eea
where $\dd{}\Omega_{d-1}^2$ is the line element of a $d-1$-dimensional unit sphere. The event horizon is located at the outermost radius $\rho=\rho_h$ at which  $f(\rho_h)=0$. 
\begin{figure}
\centering
  \includegraphics[width=0.45\linewidth]{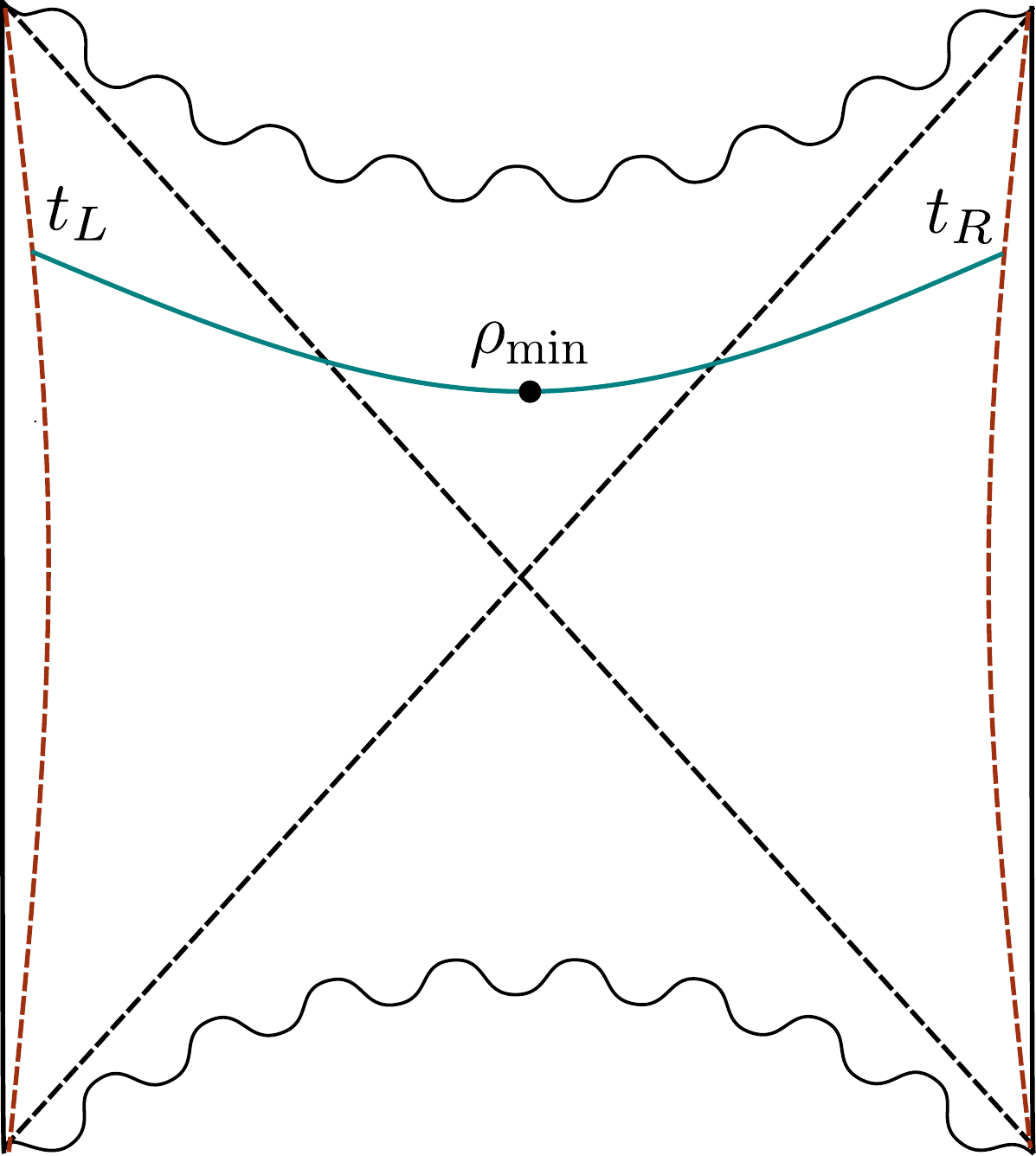}
\caption{\small Penrose diagram of an eternal AdS-Schwarzschild black hole in $(d+1)$ dimensions with a spacelike extremal volume surface anchored to fixed area cutoff surfaces at boundary times $t_{R,L}$. The turning point, $\rho=\rho_{\min}$, is in the black hole interior.}
\label{extremalcvfig}
\end{figure}
Now consider a spherically symmetric codimension-1 surface anchored to a cutoff surface in the asymptotic region (shown in Figure~\ref{extremalcvfig} for a two-sided AdS-Schwarzschild black hole).  We will consider the symmetric case where the surface is anchored at boundary times $t_R=t_L\equiv \tau$. The volume of the surface can be expressed as the following integral,
\be
{\cal V} = \Omega_{d-1}\int \dd{\lambda}\, \rho^{d-1}\sqrt{-f(\rho) \dot{t}^2+\dot{\rho}^2/f(\rho)}\, , \label{volumeeq}
\ee 
where $\Omega_{d-1}$ is the volume of the $d-1$-dimensional unit sphere and $\{t(\lambda),\rho(\lambda)\}$ parametrises
the longitudinal coordinates of the surface.
The derivatives in \eqref{volumeeq} are with respect to $\lambda$. Up to constant multiplicative factors, the holographic volume complexity of the black hole at boundary time $\tau$ is obtained by extremizing the volume functional in \eqref{volumeeq} \cite{Stanford:2014jda}, 
\bea
C_{V}(\tau) = \frac{\mathcal{V}_\textrm{ext}(\tau)}{G_{d+1}L_0}\label{cveq1}\,.
\eea
Here, $L_0$ is a characteristic length scale of the black hole and $G_{d+1}$ is Newton's constant in $d+1$ dimensions. In the case of an static black hole in AdS, we choose $L_0$ to be the AdS curvature scale $L$ while for a Schwarzschild or Reissner-Nordstr\"{o}m black hole in asymptotically flat spacetime we would instead use the radial coordinate at the event horizon.

The extremal surface has a turning point at $\rho =\rho_{\min}$ in the interior of the black hole. 
A straightforward calculation yields the following expression for the growth rate of the complexity 
(see {\it e.g.} \cite{Carmi:2017jqz} and Section~\ref{2dsection} below),
\bea
 \frac{\dd{}C_{V}}{\dd{}\tau} = \frac{2\Omega_{d-1}}{G_{d+1}L_0}\sqrt{-f(\rho_{\text{min}})}\rho^{d-1}_{\text{min}}\,.\label{cvgrowthrate}
\eea
At late times, the extremal volume surfaces approach a constant radius slice in the interior of the black hole. We will refer to this limiting surface as the \textit{accumulation surface}. The radial location of the accumulation surface, which we will denote by $\rho_{a}$, is given by \cite{Carmi:2017jqz,Stanford:2014jda,Fischetti:2014zja},
\bea
\frac{\dd{}}{\dd{\rho}}\left(f(\rho)\rho^{2d-2}\right)\Big\vert_{\rho=\rho_a} =0\,.\label{accumulationhigherdlocation}
\eea
This gives us a constant late-time growth rate  \cite{Carmi:2017jqz,Stanford:2014jda},
\bea
\frac{\dd{}C_V}{\dd{}\tau} = \frac{2\Omega_{d-1}}{G_{d+1}L_0}\sqrt{-f(\rho_{a})}\rho^{d-1}_{a} \,, \label{volgrowthrateeq1}
\eea
and in classical gravity this growth persists forever. Furthermore, 
if the characteristic length scale $L_0$ in the definition of $C_{V}$ in equation \eqref{cveq1} is chosen so as to be 
consistent with the more refined Complexity=Action proposal \cite{Brown:2015bva}, then 
the late-time rate of growth of the holographic complexity (in units of Rindler time) of a generic spherically symmetric static black hole is 
found to be proportional to the number of degrees of freedom carried by the 
black hole \cite{Brown:2015lvg},
\bea
\frac{1}{T}\frac{\dd{}C_V}{\dd{}\tau}  \propto S \,,
\eea
where $S$ and $T$ are the black hole entropy and temperature.

\section{Complexity = Length}
\label{2dsection}
The volume problem of the previous section reduces in a natural way to finding an extremal length in an auxiliary two-dimensional spacetime.  This way of phrasing the problem, which was already noticed in \cite{Stanford:2014jda}, will be particularly useful when discussing the saturation of complexity.
To see this, we perform a spherical reduction of the $d+1$-dimensional metric to two dimensions,
\bea
\dd s_{d+1}^2= \Phi^{-2} g_{\alpha\beta}\dd x^{\alpha} \dd x^{\beta}+\Phi^\frac{2}{d-1}\ell_p^2 \dd \Omega_{d-1}^2\,,\label{reductionansatzvol1}
\eea
where $\ell_p$ is the Planck length of the higher-dimensional theory, rescaled by a power of 
$\Omega_{d-1}$ to simplify some formulas below,
\be
\ell_p^{d-1} \equiv \f{G_{d+1}}{\Omega_{d-1}}\, .
\label{GNabsorbed}
\ee
The dynamical variables of the reduced theory, {\it i.e.} the two-dimensional metric $g_{\alpha\beta}$ and the dilaton field $\Phi$, are assumed to only depend on the two-dimensional coordinates $x^\alpha$. In \eqref{reductionansatzvol1} we have chosen a Weyl frame for the two-dimensional metric so that the volume measured in the $d+1$-dimensional metric is equivalent to a geodesic length measured in the 2-dimensional metric. Indeed, by expressing the black hole background in \eqref{staticbhmetric} in terms of the  two-dimensional variables, 
\be
-f(\rho)\, \dd {t}^2+\frac{\dd {\rho}^2}{f(\rho)} = \Phi^{-2}g_{\alpha\beta}\dd {x}^{\alpha}\,\dd {x}^{\beta}\, , \quad \quad \quad \quad  \rho^{d-1}=\Phi \,\ell_p^{d-1} \, .
\label{bhin2d}
\ee 
and inserting into the volume integral \eqref{volumeeq} we obtain
\be
{\cal V} =  \ell_p^{d-1} \Omega_{d-1} \, \ell\,,
\ee
where $\ell$ is the geodesic length computed using the two-dimensional metric,
\be
\ell\equiv\int \dd{\lambda} \sqrt{g_{\alpha\beta}\dot{x}^{\alpha}\dot{x}^{\beta}}\,.
\ee
The computation of the holographic volume complexity thus reduces to finding an extremal length in the two-dimensional geometry specified by \eqref{reductionansatzvol1},
\be\label{ComplexityLength}
C_V = \f{{\cal V_\textrm{ext}}}{G_{d+1}L_0} = \f{\ell_\textrm{ext}}{L_0}\,,
\ee
and in this Weyl frame there is no explicit dependence on the dilaton field. 

To proceed further, we identify two-dimensional coordinate time with its higher-dimensional counterpart and adopt
a static ansatz for the two-dimensional metric and dilaton field,
\be
g_{\alpha\beta}\dd x^{\alpha} \dd x^{\beta}=-\xi(r)\,\dd t^2+\frac{\dd r^2}{\xi(r)}\,,\qquad\qquad \Phi=\Phi(r)\,.
\label{2danzatz}
\ee
A static black hole metric of the form \eqref{staticbhmetric} translates into,
\be
\label{translation}
\xi(r)=\left(\frac{\rho(r)}{\ell_p}\right)^{2(d-1)}f\big(\rho(r)\big)\,,\qquad \Phi(r)=\left(\frac{\rho(r)}{\ell_p}\right)^{d-1}\,,
\ee
where 
\be
\label{varrel}
\frac{\dd\rho}{\dd r}=\left(\frac{\rho}{\ell_p}\right)^{-2(d-1)}\,.
\ee
We can now follow a standard route to obtain the proper length of a spacelike geodesic (see \cite{Carmi:2017jqz} for the 
corresponding computation in $d+1$ dimensions before dimensional reduction).
The first step is to switch to infalling Eddington-Finkelstein coordinates,
\be\label{2DEFmetric}
\dd s^2 = - \xi(r)\dd v^2+2 \dd r \dd v \qquad \text{where}\qquad \dd v = \dd t + \f{1}{\xi(r)} \dd r\,,
\ee
and parametrize the spacelike geodesic $x^\mu(\lambda)$ so as to have unit tangent vector everywhere, 
\be\label{unormalization}
u^\mu = \dot{x}^\mu \,,\qquad u\cdot u = 1\,.
\ee
Since $\partial_v$ is a Killing vector we have a conserved quantity,
\be\label{pdefinition}
p \equiv -u_v= \xi(r)\, \dot{v}-\dot{r} \,,
\ee
which can be combined with the tangent vector normalization in \eqref{unormalization} to obtain 
\be\label{rdotdiffeq}
\dot r^2-\xi(r)= p^2\,,
\ee
where the dot denotes a derivative with respect to $\lambda$.

\begin{figure}[h!]
\centering
  \includegraphics[width=0.65\linewidth]{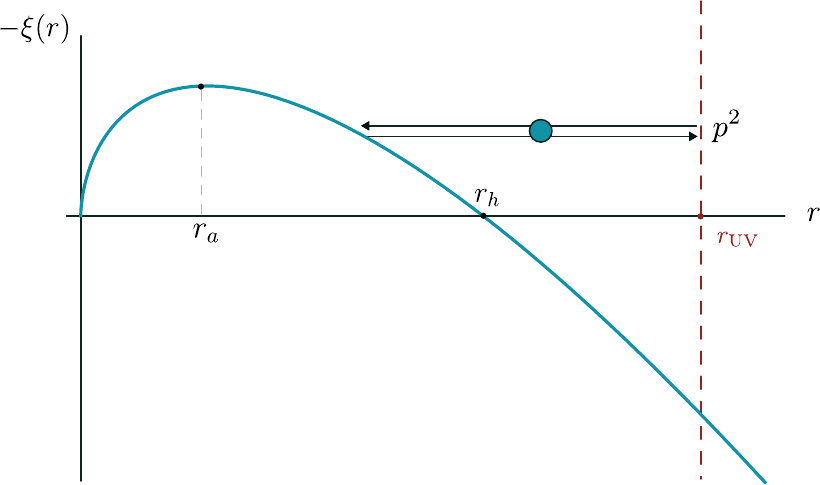}
\caption{\small One-dimensional effective potential $V_{\text{eff}} = -\xi(r)$ for an AdS-Schwarzschild black hole.}
\label{effpotentialfig}
\end{figure}

At this point it is useful to introduce a simple classical mechanics analogy where equation \eqref{rdotdiffeq} expresses energy conservation for a massive particle moving in one dimension with the affine parameter $\lambda$ playing the role of time and $r(\lambda)$ the particle location.  The effective potential is given by $V_\text{eff}=-\xi(r)$ and $p^2$ is the total energy. 
Figure~\ref{effpotentialfig} shows the effective potential obtained by applying the spherical reduction \eqref{reductionansatzvol1} to a higher-dimensional eternal AdS-Schwarzschild black hole (see Appendix~\ref{flatspacesection} for explicit formulas). The effective potential vanishes at the event horizon and has a smooth maximum at 
$r=r_a$ inside the black hole region. As we will see below, the behaviour of  
extremal volume surfaces at late boundary times in the original higher-dimensional theory is controlled by 
the effective potential near its maximum. In particular, $r=r_a$ is the radial location of the accumulation surface. 

These features are not unique to
AdS-Schwarzschild black holes. In fact, every spherically symmetric static black hole solution in $d\geq 2$ that we have considered gives rise to an effective potential that vanishes at the event horizon and has a positive maximum inside
the black hole. Several examples are worked out in Appendix~\ref{flatspacesection}, including AdS-Schwarzschild, Schwarzschild, dS-Schwarzschild, and Reissner-Nordstr\"{o}m black holes. In the charged black hole case, 
the accumulation surface is located in the region between the outer and inner horizons.

Our auxiliary two-dimensional spacetime inherits the Penrose diagram in Figure~\ref{extremalcvfig} (with the
relabelling $\rho_\text{min}\rightarrow r_\text{min}$). 
We are interested in spacelike geodesics that extend between timelike anchor curves in the left and right quadrants. 
In the classical mechanics analogy, such a geodesic corresponds to a particle with total energy in the range 
$0<p^2<-\xi(r_a)$ that travels inwards from 
$r=r_\text{UV}$ until it comes to a turning point inside the black hole at $r=r_\text{min}>r_a$, where 
$p^2 = -\xi(r_\text{min})$, and then returns to $r=r_\text{UV}$. 
At the turning point we have $u^r=\dot{r}=0$ but $u^v=\dot{v}\neq 0$ and it follows that the return journey of the 
particle corresponds to continuing the spacelike geodesic through the black hole interior to emerge on the 
opposite side of the Penrose diagram.\footnote{The sign of the conserved charge $p$ in \eqref{pdefinition} determines 
whether the geodesic is parametrised left-to-right or right-to-left in Figure~\ref{extremalcvfig}.} 
The proper length of the spacelike geodesic is given by the total time it takes the particle to make the return trip 
from $r_\text{UV}$ to the turning point and back again, which in turn depends on the particle energy $p^2$.
As the energy approaches the maximum of the effective potential, the particle spends more and more time moving
very slowly near its turning point, and in the limit $p^2\rightarrow -\xi(r_a)$ the geodesic length diverges.
A particle with energy $p^2>-\xi(r_a)$ corresponds to a spacelike geodesic that runs into the singularity at $r=0$ 
and does not connect the two asymptotic regions.

Using equation \eqref{rdotdiffeq} the geodesic length can be expressed as a radial integral,
\be\label{lengthdef}
\ell = \int \dd \lambda = 2\int_{r_\text{min}}^{r_\text{UV}} \f{\dd r}{\sqrt{p^2+\xi(r))}}\,.
\ee
The factor of two reflects the fact that the return leg of the particle trajectory takes the same amount of time as the infall. 
Placing the anchor curves at a finite radial distance regulates the otherwise divergent integral. The precise value of  
$r_\text{UV}$ is not important as we are usually only interested in the derivative of the geodesic length with respect to 
the boundary time $\tau$. The integral in \eqref{lengthdef} depends on the boundary time via the conserved 
charge $p$ and also through $r_\text{min}$, the location of the turning point, which enters both in the integrand and 
as the lower limit of integration. The 
integrand diverges as $r\rightarrow r_\text{min}$, and so some intermediate steps are in order before taking the
$\tau$ derivative.  The infalling Eddington-Finkelstein time interval along the geodesic, from the turning point 
to where it intersects the anchor curve, can be written as another radial integral,
\be\label{timedef}
v_\text{UV} - v_\text{min}
= \int_{r_\text{min}}^{r_\text{UV}}\dd r\, \frac{1}{\xi(r)}\Big(1+\f{p}{\sqrt{p^2+ \xi(r)}}\Big)\,,
\ee
where we have used equations \eqref{pdefinition}, \eqref{rdotdiffeq} and parametrised the 
geodesic left-to-right in Figure~\ref{extremalcvfig} so that $p<0$.
This expression can be combined with \eqref{lengthdef} to obtain
\be
\label{elleq}
\ell + 2\,p\, (v_\text{UV} - v_\text{min})=2 \int_{r_\text{min}}^{r_\text{UV}} \dd r\, \frac{1}{\xi(r)}\Big(\sqrt{p^2+ \xi(r)}+p\Big)\,.
\ee
Here the integrand is well defined as $r\rightarrow r_\text{min}$ and, since we have $p<0$, the integrand
is also well behaved across the horizon at $r=r_h$. Differentiating the terms in \eqref{elleq}
with respect to $\tau$ results in  
\be
\label{tauderivatives}
\frac{\dd \ell}{\dd\tau}
+2p\,\left(\frac{\dd v_\text{UV}}{\dd \tau}-\frac{\dd v_\text{min}}{\dd \tau}\right)
=-\frac{2p}{\xi(r_\text{min})}\frac{\dd r_\text{min}}{\dd \tau}\,,
\ee
where we have used \eqref{timedef} to cancel the terms that involve a factor of $\frac{\dd p}{\dd \tau}$.
The Eddington-Finkelstein time $v$ is related to Schwarzschild time $\tau$ through $v = \tau + r^*$, 
where $r^* = \int \dd r/\xi(r)$, and since the radial position of the anchor curve $r=r_\text{UV}$ is independent 
of $\tau$, it follows that
\be
\label{tauderivatives2}
\frac{\dd v_\text{UV}}{\dd\tau}=1\,.
\ee
Furthermore, for the geodesics that we consider, the interior Schwarzschild time at the turning point 
is $\tau_\text{min}=0$ for all $\tau$ and we obtain
\be
\label{tauderivatives3}
\frac{\dd v_\text{min}}{\dd \tau}=\frac{\dd r^*_\text{min}}{\dd\tau}=\frac{1}{\xi(r_\text{min})}\frac{\dd r_\text{min}}{\dd\tau}\,.
\ee
Inserting \eqref{tauderivatives2} and \eqref{tauderivatives3} into \eqref{tauderivatives} immediately 
leads to the following rather compact expression for the rate of growth of the geodesic length,
\be\label{growthrate}
\f{\dd \ell}{\dd \tau} = -2 p(\tau)\,,
\ee
in agreement with \cite{Carmi:2017jqz}. 
\begin{figure}
\centering
  \includegraphics[width=0.65\linewidth]{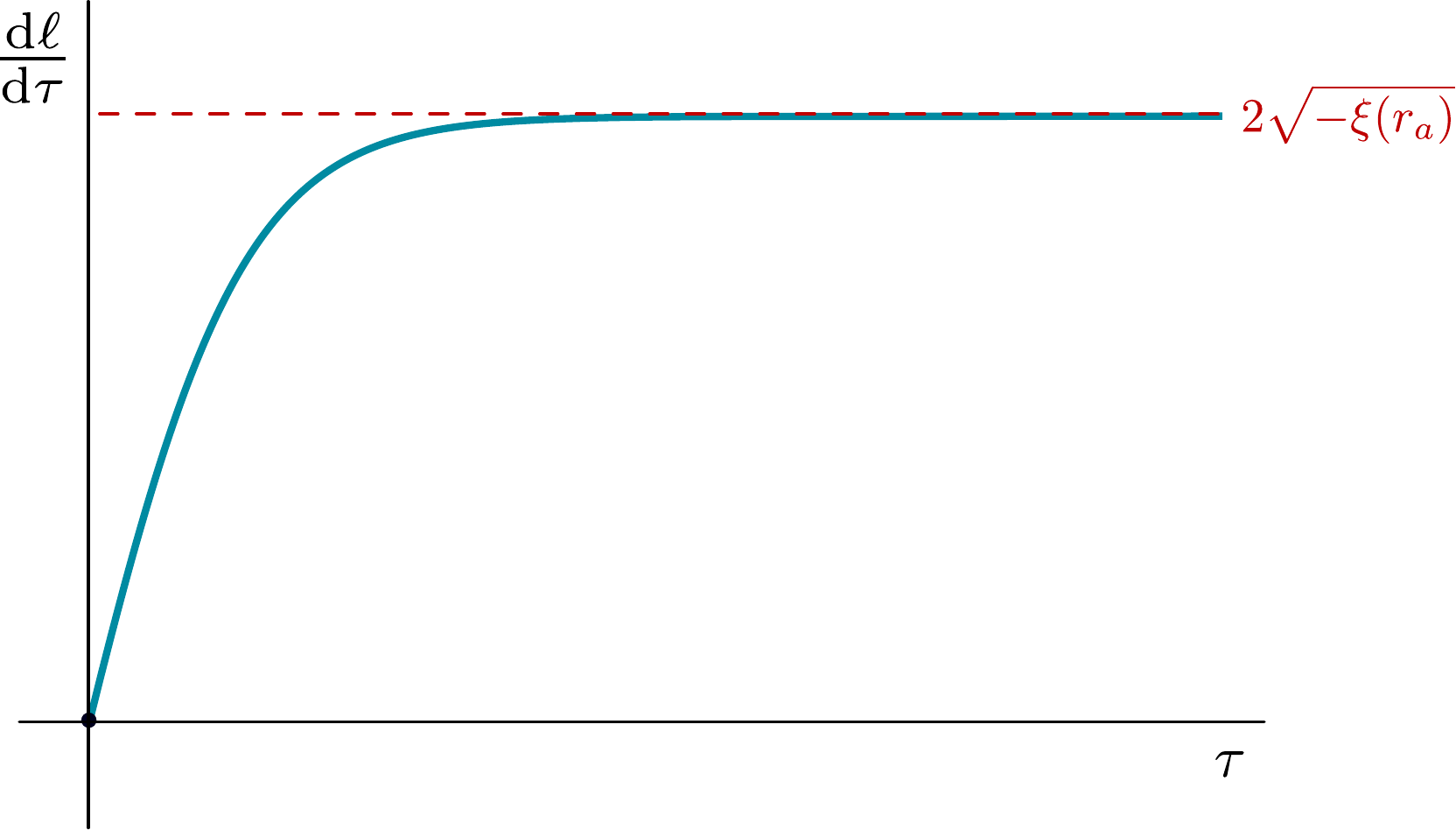}
\caption{\small 
Rate of growth of the proper length of a spacelike geodesic anchored symmetrically in the two asymptotic regions as
a function of the boundary anchor time $\tau$. The growth rate increases from zero at $\tau=0$ until it saturates as the energy of the corresponding classical particle approaches the critical point of the effective potential.}
\label{effpotentialfig1}
\end{figure}
The $\tau$ dependence of $p$, which is implicitly determined via \eqref{lengthdef} and \eqref{timedef}, can
only be solved for numerically in general but key features (sketched in Figure~\ref{effpotentialfig1})
are easily inferred from the classical mechanics analogy. 
First of all, at $\tau=0$ the spacelike geodesic is a
straight horizontal line that intersects the horizon at the bifurcation point in the Penrose diagram in 
Figure~\ref{extremalcvfig}. The corresponding classical particle turns around at $r_\text{min}=r_h$ and 
from the effective potential in Figure~\ref{effpotentialfig} we read off $p(0)=0$. Then, as $\tau$ is increased
from zero, the turning point moves into the black hole interior and the particle energy $p^2$ is a monotonically
growing function of $\tau$. At late boundary times, the energy approaches the local maximum of the
effective potential at $r=r_a$. This leads to a constant growth rate for the geodesic length,
\be
\label{laterate}
\f{\dd \ell}{\dd \tau} \rightarrow 2 \sqrt{-\xi(r_a)}\,,
\ee
which in turn translates into equation \eqref{cvgrowthrate} for the late-time rate of growth of the volume complexity
in the original higher-dimensional variables. 

In this section, we have shown how to map the holographic volume complexity of a spherically 
symmetric black hole in any number of spacetime dimensions onto the problem of determining the proper
length of spacelike geodesics in a two-dimensional geometry. In the following section, we will see how the
two-dimensional problem simplifies at late times to the calculation of geodesic length in JT gravity, 
which has been extensively studied in the literature (see {\it e.g.} 
\cite{Brown:2018bms,Iliesiu:2021ari,Alishahiha:2022kzc,Iliesiu:2024cnh,Stanford:2022fdt}).

\section{Emergent JT Gravity Description}
\label{JTgravitysec}

\begin{figure}
\centering
  \includegraphics[width=0.65\linewidth]{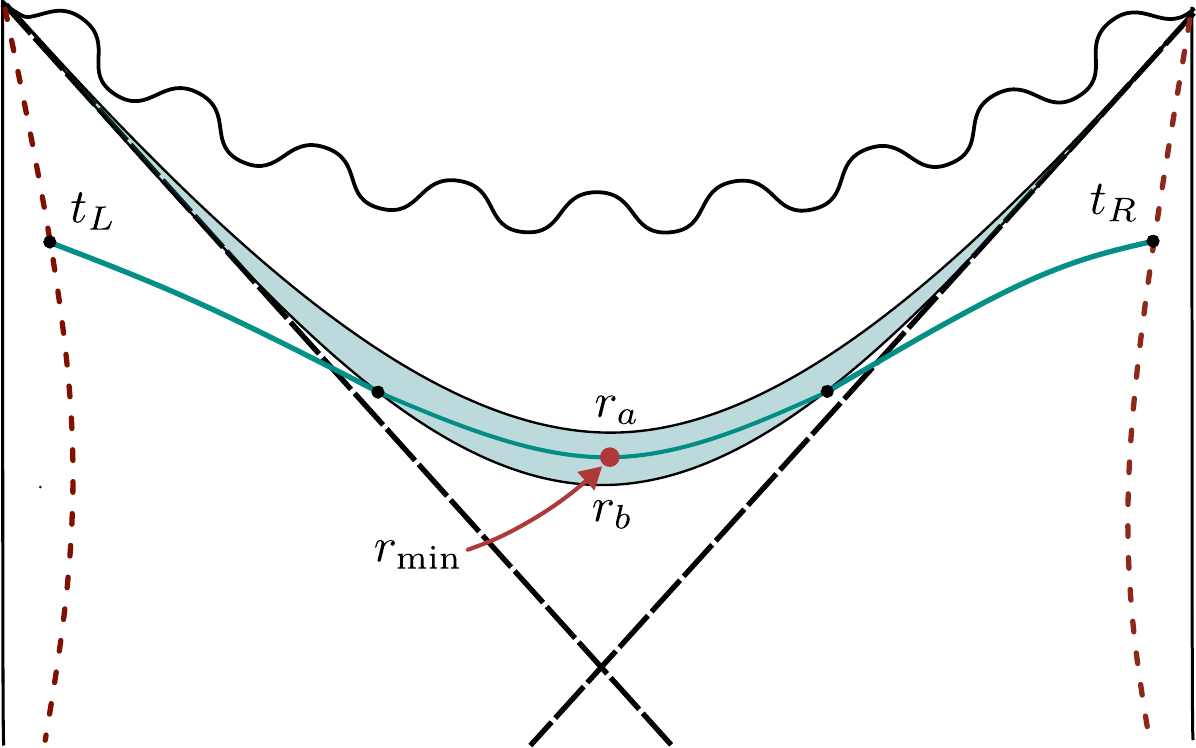}
\caption{\small The region shaded in blue is bounded by the accumulation surface $r=r_a$ and another constant radial 
surface defined by $r=r_b>r_a$ in the interior of the black hole. The parts of the spacelike geodesic that lie outside this 
region have constant length at late boundary anchor times.}
\label{extremalcvfig1}
\end{figure}

In the previous section, we re-expressed holographic volume complexity in terms of the proper length of spacelike 
geodesics in a two-dimensional dimensional auxiliary spacetime. In this section, we begin with a simple but important 
observation about the length of these geodesics. 

\subsection{A late-time JT limit}
\label{maincalsec}

Consider a fixed radial curve defined by $r=r_b$, with $r_a<r_b<r_h$, in the region of the auxiliary spacetime that corresponds to the interior of the black hole, as in Figure~\ref{extremalcvfig1}. At sufficiently late times, our boundary-anchored spacelike geodesic crosses this curve, {\it i.e.} $r_a<r_\text{min}<r_b$. We can then write the length \eqref{lengthdef} of the spacelike geodesic as a sum of two terms coming from `inside' and `outside' the $r=r_b$ curve, 
\bea
\begin{aligned}
\ell  &= 2\int_{r_\text{min}}^{r_b} \f{\dd r}{\sqrt{\xi-\xi(r_\text{min})}}\,+2\int_{r_b}^{r_\text{UV}} \f{\dd r}{\sqrt{\xi-\xi(r_\text{min})}}\,\\
&\equiv \tilde{\ell}+2\int_{r_b}^{r_\text{UV}} \f{\dd r}{\sqrt{\xi-\xi(r_\text{min})}}\,,
\end{aligned}
\eea
where we have defined $\tilde\ell$ as the length of the section of the geodesic that is inside $r=r_b$.
Taking a derivative with respect to $\tau$, we get
\bea
\label{growthestimate}
\frac{\dd{\left(\ell- \tilde{\ell}\right)}}{\dd{\tau}}=\frac{\dd{\xi(r_\text{min})}}{\dd{\tau}}\int_{r_b}^{r_\text{UV}} \f{\dd r}{\left(\xi-\xi(r_\text{min})\right)^{3/2}}\,
\eea
At late times, the turning point approaches the constant radial location of the accumulation surface
so that $\frac{\dd{\xi(r_\text{min})}}{\dd{\tau}}\rightarrow 0$. Since $r_b>r_a$, the integral in \eqref{growthestimate} 
remains finite in the limit and this implies that the right hand side vanishes as $r_\text{min}\to r_a$. 
Integrating at late time thus gives 
\bea
\ell(\tau) \approx \tilde{\ell}(\tau)+\ell_0 \,,
\label{truncatedleneq1}
\eea
where $\ell_0$ is a time independent constant. In other words, the late-time growth of the geodesic length 
comes from the region near the accumulation surface (shaded in blue in Figure~\ref{extremalcvfig1}). 
This approximation becomes better and better as time passes.

A second observation is that the auxiliary two-dimensional spacetime has negative curvature at the 
accumulation surface,
\be
R\,\vert_{r=r_a} = -\xi''(r_a) < 0\,.
\ee
This is immediately apparent from the shape of the effective potential in Figure~\ref{effpotentialfig}. 
Since we are interested only in the late-time growth rate of the geodesic length, we can restrict our attention 
to the region near the accumulation surface. Introducing a new radial coordinate $\eta$, defined by
\bea
r=r_a +\eta\,, \label{volaccexpansioneq1}
\eea
with $\eta\ll r_a$, and expanding in powers of $\eta$ leads to 
\bea
\xi(r) = \xi(r_a)+\xi^{\prime}(r_a)\eta+\frac{\xi^{\prime\prime}(r_a)}{2}\eta^2 +O(\eta^3) \,.\label{metrictayloreq}
\eea
Since $r_a$ is a critical point $\xi$, the linear term in the $\eta$ expansion vanishes. Truncating the expression at the quadratic order, and plugging it back into the metric, leads to
\begin{equation}
\dd{s}^2=-\Big( \xi(r_a)+\frac{\xi^{\prime\prime}(r_a)}{2}\eta^2\Big) \dd{t}^2+\frac{\dd{\eta}^2}{\Big( \xi(r_a)+\frac{\xi^{\prime\prime}(r_a)}{2}\eta^2\Big)}\,.\label{latetimevolmetric}
\end{equation}
The accumulation surface lies inside the horizon and hence $\xi(r_a)<0$. Therefore, we can identify 
\eqref{latetimevolmetric} with an AdS$_2$ black hole metric,
\bea
\dd{s}^2=-\left(\frac{\eta^2}{L_2^2}-\mu\right)\dd{t}^2+\frac{\dd{\eta}^2}{\left(\frac{\eta^2}{L_2^2}-\mu\right)}\,, \label{AdS2metriceq}
\eea
where $L_2$ is an \emph{emergent} AdS$_2$ length scale and $\mu$ a mass parameter, 
\bea
L_2^2 = \frac{2}{\xi^{\prime\prime}(r_a)}, \quad \quad \mu = -\xi(r_a)\,. \label{ads2lengdefeq1}
\eea
The horizon of this emergent black hole is located at $\eta_{h}= \sqrt{\mu}L_2$ and its temperature is given by
\bea
T_2 = \frac{\eta_h}{2\pi L^2_2} \equiv \frac{1}{\beta_2}\,.   \label{betaeq}
\eea
If the $r=r_b$ curve in Figure~\ref{extremalcvfig1} is placed near the accumulation surface, {\it i.e.}
\bea
r_b-r_a\ll r_a\,, \label{vicinityeq}
\eea
then the AdS$_2$ black hole metric \eqref{AdS2metriceq} is valid over 
the entire spacetime region where the late-time complexity growth takes place.\footnote{The fact that we find 
an emergent AdS$_2$ geometry when expanding around the accumulation surface is perhaps not surprising. 
Any two-dimensional metric with a timelike Killing symmetry can be brought to the form \eqref{2danzatz} and 
an expansion to quadratic order always leads to a spacetime metric with constant curvature. What is important 
about the expansion around the accumulation surface is that it leads to a spacetime metric with constant 
\emph{negative} curvature.}

\begin{figure}
  \centering
    \includegraphics[width=0.45\linewidth]{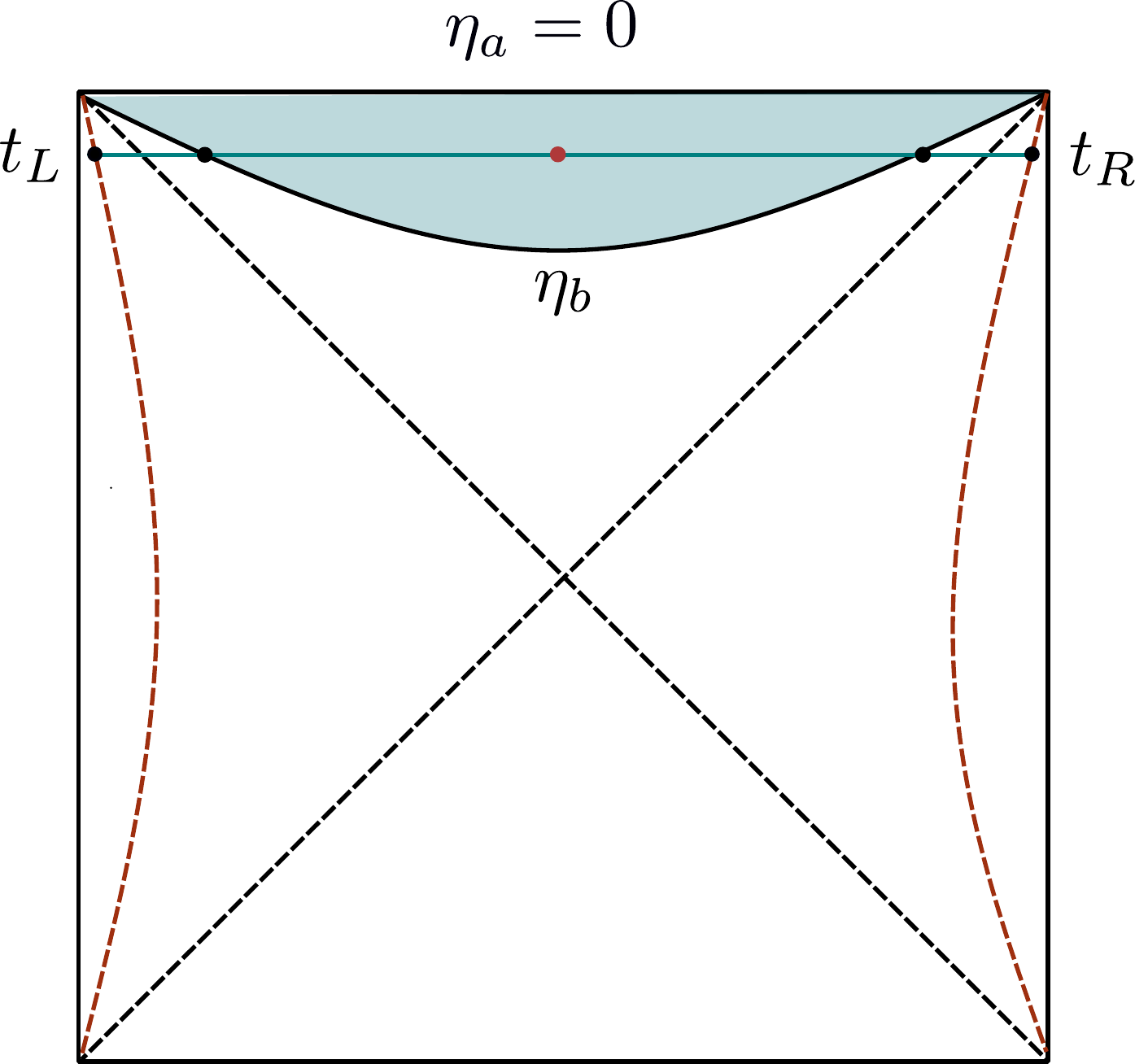}
  \caption{\small Penrose diagram for the AdS$_2$ black hole \eqref{latetimevolmetric}. 
  The length of the geodesic inside the 
  region enclosed by the $\eta=\eta_b$ and $\eta=0$ curves (shaded in blue) is denoted by $\tilde{\ell}_\text{JT}$ 
  in the main text, while the length of the full geodesic anchored to a constant $\eta$ surface in the asymptotic 
  AdS$_2$ region is denoted by $\ell_\text{JT}$. 
  }
  \label{ads2penrosefig}
  \end{figure}
The Penrose diagram for the AdS$_2$ black hole is shown in Figure~\ref{ads2penrosefig}. 
The shaded region, between the $\eta=0$ and $\eta=\eta_b\equiv r_b-r_a$ curves, corresponds to the 
similarly shaded region in Figure~\ref{extremalcvfig1}. The condition \eqref{vicinityeq} 
ensures that the curve $\eta=\eta_b$ lies well inside the horizon of the AdS$_2$ black hole. 
We are interested in spacelike geodesics that are symmetrically anchored in the asymptotic  
AdS$_2$ region at boundary times $t_L=t_R=\tau$. Such geodesics are straight horizontal curves of constant 
global time in the AdS$_2$ Penrose diagram \cite{Brown:2018bms} and at sufficiently late boundary times 
they will intersect the shaded region in Figure~\ref{ads2penrosefig}.

The analysis of spacelike geodesics in Section~\ref{2dsection}, including the classical mechanics analogy,
can be applied to the emergent AdS$_2$ metric \eqref{AdS2metriceq}. The resulting effective potential
takes a particularly simple form,
\be
V_\text{eff}(\eta)=\mu-\frac{\eta^2}{L_2^2}\,,
\ee
with $L_2$ and $\mu$ defined in \eqref{ads2lengdefeq1}.
This corresponds to zooming in on the critical point of the effective potential at $r=r_a$ in 
Figure~\ref{effpotentialfig} and expanding to quadratic order in $\eta=r-r_a$. 
At late boundary times, our spacelike geodesics can be divided into a part that passes through
the shaded region inside 
the curve $\eta=\eta_b$ in Figure~\ref{ads2penrosefig} and external parts that attach onto the anchor 
curves in the asymptotic AdS$_2$ regions. Accordingly, the geodesic length consists of two terms,
\be
\ell_\text{JT}(\tau)=\tilde\ell_\text{JT}(\tau)+\ell_\text{JT}^{\,0} \,,
\label{fullJTlength}
\ee
where the JT subscript indicates that this is in fact the length
of a boundary anchored spacelike geodesic in an 
emergent two-dimensional Jackiw-Teitelboim dilaton gravity theory that we will elaborate on below.

The same line of reasoning that led to \eqref{truncatedleneq1} implies that the
outside length $\ell_\text{JT}^{\,0}$ approaches a time independent constant at late times. 
Since the emergent AdS$_2$ metric only applies near the accumulation surface, and the location of the AdS$_2$
anchor curve differs from the original anchor curve, the full geodesic length at a given late boundary time 
will differ between  \eqref{truncatedleneq1} and \eqref{fullJTlength}. 
However, if $r_b$ (or equivalently $\eta_b$) is chosen so that the shaded region in Figure~\ref{extremalcvfig1} 
is well described by the emergent AdS$_2$ metric, then the {\it time dependent\/} part of the geodesic length 
is the same in both descriptions at late times, 
\be
\tilde\ell(\tau)=\tilde\ell_\text{JT}(\tau)\,,
\ee
It follows that the late-time volume complexity can be expressed as
\be
C_V(\tau)\approx C_0+\frac{\ell_\text{JT}(\tau)}{L_0}\,, \label{volaslengeq}
\ee
where $C_0$ is a time independent constant.
The problem of determining the late-time growth of 
volume complexity of a spherically symmetric black hole in any number of spacetime dimensions thus maps 
onto the problem of determining the late-time growth of geodesic length in a two-dimensional geometry of 
constant negative curvature, {\it i.e.} two-dimensional locally anti-de Sitter spacetime. This rather general result 
is one of the main conclusions of this paper. The characteristic length scale in \eqref{ads2lengdefeq1} 
of the JT gravity theory in question is inherited from the higher-dimensional, static, spherically symmetric, 
black hole solution that we start with (see appendix \ref{flatspacesection} for some examples). 
In the following subsection we show how the action of two-dimensional JT gravity arises from spherical reduction
in this context.

\subsection{Emergent JT action}
\label{JTreduction}

Applying the spherical reduction ansatz \eqref{reductionansatzvol1} to  $d+1$-dimensional Einstein gravity 
yields a two-dimensional dilaton-gravity theory,
\be
S=\frac{1}{16 \pi}\int \dd^2 x \sqrt{-g}\Big(\Phi R^{}-\frac{d}{d-1}\frac{(\nabla\Phi)^2}{\Phi}+W(\Phi)\Big),
\label{bulkdilatonactionkineq1}
\ee
that has been analysed extensively in the past by numerous authors (see {\it e.g.} \cite{Grumiller:2007ju} for a review). 
The  form of the dilaton potential $W(\Phi)$ depends on details of the higher-dimensional theory, 
such as the value of the $d+1$-dimensional cosmological constant for instance.
The equations of motion that follow from \eqref{bulkdilatonactionkineq1} can be expressed as 
\begin{equation}
\begin{split}\label{eomdilatoneq1}
0&=\nabla_{\mu} (\Phi^{\f{d}{d-1}}\nabla_{\nu} \Phi)-\frac{1}{2} g_{\mu \nu} \nabla\cdot(\Phi^{\f{d}{d-1}}\nabla \Phi)\,, \\
0&=\nabla^2 \Phi-W(\Phi)\,, \\
0&=R+\frac{d}{(d-1)}\Big(2\frac{\nabla^2 \Phi}{\Phi}
-\frac{(\nabla\Phi)^2}{\Phi^2}\Big)+ W'(\Phi)\,.
\end{split}
\end{equation}
The first equation can be interpreted as Killing's equation for 
\be
K^\mu = \Phi^{\f{d}{d-1}}\epsilon^{\mu\nu}\nabla_{\nu}\Phi\,,
\ee
and the value of the dilaton is preserved along the Killing vector ${\cal L}_K\Phi = 0$ \cite{Banks:1990mk}. 
Our starting point is some static black hole solution of the higher-dimensional theory, which suggests we 
take $K^\mu$ to be timelike. A two-dimensional metric with a time-like Killing vector can always be put
into the form of \eqref{2danzatz} with a time-independent dilaton field.

Plugging the ansatz \eqref{2danzatz} into the equations of motion \eqref{eomdilatoneq1}, we obtain simple ODEs that  
can be explicitly solved for arbitrary dilaton potential \cite{Grumiller:2007ju}, 
\bea
\begin{aligned}
\partial_{{r}} \Phi &= e^{-Q_0}\Phi^{-\frac{d}{d-1}}\,,\\
\xi(\Phi) & = e^{Q_0}\Phi^{\frac{d}{d-1}}w(\Phi)\,,
\end{aligned}\label{metriceom1}
\eea
where the function $w(\Phi)$ is an integral involving the dilaton potential
\bea
w(\Phi) =w_0+ \int^\Phi d\tilde{\Phi} \ W(\tilde{\Phi}) \ \tilde{\Phi}^{\frac{d}{d-1}}e^{Q_0}\,,
\eea
and $Q_0$, $w_0$ are constants of integration whose values are determined by the higher-dimensional
black hole solution via \eqref{translation}. 

The accumulation surface inside the higher-dimensional black hole corresponds to a spacelike curve 
of constant dilaton field $\Phi= \Phi_a\equiv\Phi(r_a)$ in the two-dimensional spacetime of the dilaton gravity theory.
In order to zoom in on the spacetime region near the accumulation surface, we want to expand the dilaton field 
around $\Phi_a$ while also keeping track of the background value of $\nabla^{\mu}\Phi$.
Operationally, this can be achieved by introducing an auxiliary field, which, when evaluated on-shell, reproduces 
the original action \eqref{bulkdilatonactionkineq1} and its associated equations of motion. 
Concretely, we introduce a new vector field $\lambda_{\mu}$ through the action
\be
S=\frac{1}{16 \pi }\int \dd^2 x \sqrt{-g}\left(\Phi R+W(\Phi)+\frac{d}{(d-1)}\big(\lambda^2 \Phi
-2\lambda^{\mu}\nabla_{\mu}\Phi\big)\right)\,.
\label{newlambdaaction}
\ee
Varying this action with respect to $\Phi$ and $\lambda_{\mu}$, we get the following equations of motion,
\begin{eqnarray}
R+W'(\Phi)+\frac{d}{(d-1)}\big(\lambda^2 +2\nabla_{\mu}\lambda^{\mu}\big)&=&0\,,\label{newphieomeq}\\
\lambda^{\mu} - \frac{\nabla^{\mu}\Phi}{\Phi}&=&0\,.\label{lambdaeomeq}
\end{eqnarray}
Substituting \eqref{lambdaeomeq} in \eqref{newlambdaaction} and \eqref{newphieomeq}, we get back the original action \eqref{bulkdilatonactionkineq1} and its dilaton field equation in \eqref{eomdilatoneq1}. Therefore, the theories described by \eqref{newlambdaaction} and \eqref{bulkdilatonactionkineq1} are classically equivalent.

Now let us expand the fields $\Phi$ and $\lambda_{\mu}$ around some constant values,
\bea
\Phi = \Phi_0 +\phi\,, \quad \quad \lambda_{\mu} = \Lambda_{\mu}+\tilde{\lambda}_{\mu}\,,
\label{philambdaexpansions}
\eea
and evaluate the action \eqref{newlambdaaction} to \textit{linear} order in the fluctuations, dropping total derivatives, 
to obtain
\bea
\begin{aligned}
S&\simeq\frac{1}{16 \pi}\int \dd^2 x \sqrt{-g}\Big(\Phi_0 R+W(\Phi_0)+\frac{d}{(d-1)}\Lambda^2 \Phi_0\Big)\\
&+\frac{1}{16 \pi }\int \dd^2 x \sqrt{-g}\Big( \big(R+W'(\Phi_0)\big)\phi
+\frac{d}{(d-1)}\big(\Lambda^2 \phi+2\Phi_0\Lambda_{\mu} \tilde{\lambda}^{\mu}\big)\Big)\,.
\end{aligned}
\label{expandedaction}
\eea
For a static solution, the only non-vanishing component of the field $\lambda^{\mu}$ is given by
\bea
\lambda^{r} = \frac{\partial^{r}\Phi}{\Phi} = e^{-Q_0}\Phi^{\frac{-2d+1}{d-1}}\xi(\Phi) \equiv U(\Phi)\,,
\eea
where we have used equations \eqref{lambdaeomeq} and \eqref{metriceom1}. 
Inserting the expansions \eqref{philambdaexpansions} on both sides, we obtain
\bea
\Lambda^{r}+\tilde{\lambda}^{r} = U(\Phi_0)+U^{\prime}(\Phi_0)\phi\,,
\eea
which allows us to identify 
\bea
\Lambda^r = U(\Phi_0) \quad \text{and} \quad \tilde{\lambda}^{r} = U^{\prime}(\Phi_0)\phi\,.
\eea
Similarly, we find that the background vector with the index lowered is given by
\bea
\Lambda_{r} = e^{-Q_0}\Phi_0^{\frac{-2d+1}{d-1}}\, .
\eea
Substituting these expressions into the action in \eqref{expandedaction} leads to
\bea
\begin{aligned}
S&=\frac{1}{16 \pi}\int \dd^2 x \sqrt{-g}\Big(\Phi_0 R+\frac{d}{(d-1)}\Lambda^2 \Phi_0+W(\Phi_0)\Big)\\
&+\frac{1}{16 \pi }\int \dd^2 x \sqrt{-g}\Big(R+W'(\Phi_0)
+\frac{d}{d-1}\big(\Lambda^2 +2\Lambda_{r} U^{\prime}(\Phi_0)\Phi_0\big)\Big)\,\phi+ O(\phi^2)\,.
\end{aligned}\label{newlambdaaction1}
\eea
Finally, if the background $\Phi_0$ is chosen to have the value of the dilaton at the accumulation surface, $\Phi_0=\Phi_a$, 
the second and third term in the first line of \eqref{newlambdaaction1} cancel each other out, and the action reduces to a simple form,
\bea
\begin{aligned}
S=\frac{\Phi_a}{16\pi} \int \dd^2 x \sqrt{-g} R
+\frac{1}{16\pi}\int \dd^2 x \sqrt{-g}\Big(R+\frac{2}{L_2^2}\Big) \,\phi + O(\phi^2)\, .
\end{aligned}\label{JTaction}
\eea
As expected, the action \eqref{JTaction} governing the dynamics near the accumulation surface, takes the form 
of JT gravity \cite{Jackiw:1984je,Teitelboim:1983ux} with the same AdS$_2$ length scale $L_2$ that we 
encountered in the JT black hole metric in \eqref{AdS2metriceq}. 
The coefficient in front of the topological term in \eqref{JTaction} is 
given by the area of the transverse $d-1$-sphere at the accumulation surface inside the higher-dimensional
black hole rather than its horizon area but the two areas only differ by an order-one multiplicative
constant that depends on the higher-dimensional black hole under consideration.

\section{Saturation of Black Hole Complexity}
\label{saturationfromJTsection}
In section \ref{JTgravitysec}, we saw how the calculation of the classical volume of extremal surfaces 
at late times in various static black hole solutions reduces to calculating a geodesic length in 
an emergent two-dimensional JT gravity theory. With this equivalence in hand, we can take advantage 
of an existing quantum mechanical description of JT gravity to arrive at a non-perturbative definition 
of complexity for static spherically symmetric black holes in any number of spacetime dimensions. 
Building on earlier work relating the partition function of JT gravity to matrix integrals 
\cite{Saad:2019lba,Saad:2019pqd},  a quantum mechanical definition of geodesic length in JT gravity 
was given in \cite{Iliesiu:2021ari} and demonstrated to saturate on extremely long time scales.
If we adopt the same quantum definition of geodesic length in our emergent JT gravity description
of black hole complexity in higher-dimensional gravity, then it immediately follows that holographic 
volume complexity saturates on extremely long time scales in the higher-dimensional case as well.
In our setup, the parameters of the JT gravity theory are inherited from the higher-dimensional 
black hole under consideration, and thus the two-dimensional results of \cite{Iliesiu:2021ari} can be 
adapted to predict the time-scale on which the higher-dimensional black hole complexity saturates.
In the remainder of this section, we summarise the results of \cite{Brown:2018bms,Iliesiu:2021ari} 
that are relevant to the present paper and then translate those results into saturation time-scales for
higher-dimensional black hole complexity via the emergent JT description. 

\subsection{Length of geodesics in JT gravity}
\label{JTgravityreviewsec}
In a two-dimensional theory, there is only one spatial dimension. The volume complexity of a 
JT black hole is therefore given by the length of a spacelike geodesic (divided by a fixed characteristic length scale), 
with the spacelike geodesic anchored to timelike boundary curves in the asymptotic AdS$_2$ regions on either
side of the two-dimensional Penrose diagram in Figure~\ref{ads2penrosefig}.
This problem was considered in \cite{Brown:2018bms}, where the geodesic length was found to grow 
linearly with the boundary anchor time, giving the expected growth rate for JT black hole complexity,
\be
\frac{\dd{}C_V}{\dd{}\tau}  \propto S\,T \,.
\ee
Here $S$ and $T$ are the Bekenstein-Hawking entropy and temperature of the $3+1$-dimensional 
near-extremal Reissner-Nordstr\"{o}m black hole, whose near-horizon region is described by
the effective JT gravity theory in \cite{Brown:2018bms}. At the classical level, this linear growth of complexity
continues forever, which is in line with extremal volume calculations in higher-dimensional gravity theories, but in contrast 
with the expected saturation of quantum complexity at very late times. However, since JT gravity is more amenable
to quantisation than its higher-dimensional cousins, one can envisage a quantum representation of geodesic length
as in \cite{Iliesiu:2021ari}. We will not repeat their detailed arguments here but simply state some of the
main results. 

The starting point is the action \eqref{JTaction} for two-dimensional JT gravity (supplemented by the usual
boundary terms). Following \cite{Maldacena:2016upp,Banerjee:2021vjy}, we impose boundary conditions,
\bea
\dd s^2 = \frac{L^2_2}{\epsilon}\dd u^2\,, \quad \phi = \frac{\phi_b}{\epsilon}\,, \label{boundarycondJT}
\eea
where $u$ is the boundary time. Here $\phi_b$ has the dimensions of length and characterises the scale of 
time reparametrisation symmetry breaking. The boundary has a fixed proper length $\frac{\beta \phi_b}{\epsilon}$, 
where $\beta$ is the inverse temperature of the black hole.

The length of spacelike geodesics anchored onto the asymptotic boundaries can be \emph{non-perturbativley}
defined in JT gravity using its matrix model description \cite{Iliesiu:2021ari,Alishahiha:2022kzc}, via
\bea
\left\langle \ell (\tau)\right\rangle = \int_0^{\infty} \dd E_1 \dd  E_2\left\langle\rho\left(E_1\right) 
\rho\left(E_2\right)\right\rangle e^{-\frac{\beta}{2}\left(E_1+E_2\right)} e^{-i \tau\left(E_1-E_2\right)} 
\mathcal{M}_\Delta\left(E_1, E_2\right)\,,\label{lengthint}
\eea
where $\rho(E)$ is the density of states. 
The quantity $\mathcal{M}_\Delta(E_1,E_2)$ is related to the matrix elements of a boundary operator $\mathcal{O}_\Delta$,
\bea
\mathcal{M}_\Delta(E_1,E_2) =-\lim _{\Delta \rightarrow 0} \frac{\partial}{\partial \Delta}\left|\left\langle E_1|\mathcal{O}_\Delta| E_2\right\rangle\right|^2\,, \label{matrixelementdefM}
\eea
where $\Delta$ is the scaling dimension of $\mathcal{O}_\Delta$. 
Heuristically, this amounts to defining the quantum mechanical length
in terms of a two-point correlation function of scaling operators in the boundary dual theory, that are
inserted at the boundary anchor time of the geodesic. The classical geodesic length is then reproduced 
in a semi-classical limit where the correlation function can be evaluated in a geodesic approximation.

Now we can use the fact that JT gravity is dual to a matrix model \cite{Saad:2019lba}, and that the two-point function 
of the density of states has a universal form when $E_1 \to E_2$ \cite{Saad:2019pqd,mehta2004random},
\be
 \left\langle\rho(E) \rho(E^{\prime})\right\rangle =\rho_0(E) \rho_0(E^{\prime})-\frac{\sin ^2\left(\pi\rho_0(\bar{E})
 \left(E-E^{\prime}\right)\right)}{\left(\pi \left(E-E^{\prime}\right)\right)^2}+\rho_0(\bar{E}) \delta\left(E-E^{\prime}\right)\,, 
 \label{doseq}
\ee
where we have defined $2\bar{E} = E+E^{\prime}$. Here $\rho_0= \e^{\frac{\Phi_a}{4}}\sinh(2\pi\sqrt{E})$ is the 
disk contribution to the density of states, with $\Phi_a$ the parameter multiplying the topological term in the 
JT gravity action in \eqref{JTaction}. The second term in the above expression is non-perturbative in 
$e^{\frac{\Phi_a}{4}}$, and it is usually referred to as the \textit{sine kernel} in the matrix model literature. 
Plugging in the explicit expressions of $\rho_0(E)$ and $\mathcal{M}_{\Delta}(E_1,E_2)$, one can evaluate 
the integral to obtain \cite{Iliesiu:2021ari},
\bea
\langle\ell(\tau)\rangle \approx \begin{cases}C_1 \tau+\ldots & \tau \ll e^{\frac{\Phi_a}{4}}\phi_b \\ C_0-\ldots & \tau 
\gg e^{\frac{\Phi_a}{4}}\phi_b\end{cases} \label{lensaturationeq}
\eea
where 
\bea
\begin{aligned}
& C_0=\frac{e^{\frac{\Phi_a}{4}}\phi_b}{6 \pi^2}\left[e^{6 \pi^2\phi_b / \beta}\left(1+\frac{16 \pi^2\phi_b}{\beta}\right)-e^{-2 \pi^2 \phi_b/ \beta}\right], \\
& C_1=2e^{-2 \pi^2 \phi_b/ \beta}\sqrt{\frac{\phi_b}{\pi \beta / 2}}+\frac{\operatorname{erf}\left(\sqrt{\frac{2\pi^2 \phi_b}{\beta}}\right)}{ \pi}\left(1+\frac{4 \pi^2 \phi_b}{\beta}\right).
\end{aligned}
\eea
The linear growth at early times can be traced back to the leading disconnected piece in \eqref{doseq}, but at very 
late times, this contribution is cancelled by another contribution that comes from the sine kernel, resulting in the 
saturation of the length.

\subsection{Late Time Behavior of Black Hole Complexity}
\label{saturationsection}
The length calculations in \cite{Iliesiu:2021ari} were performed in units where $\phi_b$ is set to 1. 
To obtain the late-time behaviour of holographic complexity in our formalism, it is necessary to obtain 
an explicit expression for $\phi_b$. Expanding the dilaton solution \eqref{metriceom1} around its value 
at the accumulation surface, we find that 
\bea
\phi = \frac{(d-1)}{(2d-1)}\frac{\Phi_a\eta}{r_a}\,. \label{boundarydilatoneq1}
\eea
The boundary conditions in JT gravity are usually specified in the Poincar\'{e} coordinate $z$, defined 
via the relation
\bea
\frac{L_2^2}{z^2} = \frac{\eta^2-\eta_h^2}{L^2_2}\,.
\eea
Choosing the cutoff surface to be at $z = \epsilon$, we can use \eqref{boundarycondJT} to 
determine the boundary value of the dilaton as follows
\bea
\phi_b = \frac{(d-1)}{(2d-1)} \frac{\Phi_a L_2^2}{r_a}\, . \label{phibvalue}
\eea
Now we can use \eqref{volaslengeq} and \eqref{lensaturationeq} to obtain the late-time behaviour of holographic complexity
\bea
C_V(\tau) \approx  \frac{\ell_0}{L_0}+\begin{cases}\mathcal{C}_1 \tau+\ldots & \tau \ll \tau_S \\ 
\mathcal{C}_0-\ldots & \tau \gg \tau_S\end{cases}\,. \label{volsaturationeq}
\eea
The saturation time $\tau_S$ can be calculated using \eqref{phibvalue} and we find that
\bea
\tau_S =e^{\frac{\Phi_a}{4}}\phi_b = \frac{(d-1)}{(2d-1)} \left(\frac{\Phi_aL_2^2}{r_a}\right)e^{\frac{\Phi_a}{4}}\,.
\label{saturationtimeeq}
\eea
Finally, we can work out $\tau_S$ and $\mathcal{C}_{0,1}$ for different black hole solutions 
by using the explicit expressions in Appendix \ref{flatspacesection} (see table \ref{table1}). To reproduce the classical growth rates in Appendix \ref{flatspacesection}, we have rescaled the boundary time $\tau$. 

For all static spherically symmetric black holes, we find that the complexity saturates at times of 
exponential order in $S$, the Bekenstein-Hawking entropy of the higher-dimensional black hole, that is,
\bea
\tau_S = e^{O\left(S\right)}  \tau_0\,. \label{saturationtimescaleeq}
\eea
where $\tau_0$ is some characteristic timescale associated to the black hole. This is precisely the expected behaviour 
of operator complexity in a chaotic system \cite{Barbon:2019wsy}.

The coefficient of $S$ in the exponent of \eqref{saturationtimescaleeq} depends on the higher-dimensional black hole 
and its spacetime dimensions (see table \ref{table1}). In the case of a near-extremal Reissner-Nordstr\"{o}m (RN) black 
hole, this coefficient is one and complexity saturates at times of the order $e^S$. As a simple consistency check, 
we note that the same timescale is obtained if we instead use the usual effective JT gravity description in the 
near-horizon region of a near-extremal RN black hole to compute $\tau_S$ directly. 
\begin{table}[t!]
\resizebox{\textwidth}{!}{\begin{tabular}{ |c||c|c|c|  }
 \hline
 \phantom{a}& AdS-Schwarzschild &Schwarzschild &Near-Extremal RN \\
 \hline
Saturation Time ($\tau_S$)  & $\exp{\left({2^{\left(\frac{1-d}{d}\right)}S+\cdots}\right)} \frac{L^2}{\rho_h}$   &$\exp{\left({\left(\frac{d}{2d-2}\right)^{\frac{d-1}{d-2}}S+\cdots}\right)}  \rho_h$&   $\exp{\left(S+\cdots\right)} \rho_+$\\
Saturation Value ($\mathcal{C}_0$)&   $\exp{\left({2^{\left(\frac{1-d}{d}\right)}S+\cdots}\right)}$  & $\exp{\left({\left(\frac{d}{2d-2}\right)^{\frac{d-1}{d-2}}S+\cdots}\right)} $   &$\frac{\delta \rho}{\rho_+}\exp{\left(S+\cdots\right)} $\\
 \hline
\end{tabular}}
\caption{\small The table lists the saturation time $\tau_s$ and the saturation value $\mathcal{C}_0$ for the volume
complexity of various $d+1$-dimensional black holes. 
Here $S$ is the Bekenstein-Hawking entropy of the black hole. The dots correspond to logarithmically subleading terms.}
\label{table1}
\end{table}

\section{Discussion}
\label{discussionsection}
In this paper we have shown that the linear growth of volume complexity for any static spherically symmetric black hole 
at late times can be understood as the linear growth of a geodesic length in an effective AdS$_2$ geometry. 
This is true for black holes in dimensions $d \ge 2$ and follows from a dimensional reduction of the extremal 
volume problem to two dimensions along with a judicious choice of a Weyl frame which transforms it to the 
calculation of a geodesic.

Since the black hole Hilbert space is expected to be finite dimensional, any quantity (like complexity) that evolves
 for a long time should eventually saturate. The fact that volume complexity of black holes does not saturate in 
 classical gravity suggests that some quantum gravity effects should take over and lead to the expected saturation. 
 As was discussed in Section~\ref{saturationfromJTsection} above, the matrix model description of JT gravity 
 provides an appealing resolution of this issue for the complexity of two-dimensional JT black holes, and by extension 
 for generic spherically symmetric black holes via the universal nature of complexity growth observed in the present paper. 
 Whether such an effect can be modelled directly in higher-dimensional quantum gravity, {\it e.g.} via some 
 non-perturbative corrections to the gravitational path integral, remains an open question (see for example \cite{Balasubramanian:2022gmo,Balasubramanian:2024yxk,Iliesiu:2024cnh,Boruch:2024kvv,Geng:2024jmm}). 

 It is intriguing that, even without a detailed understanding of the saturation mechanism in higher-dimensional gravity, 
 we can still make progress by utilising the impressive recent progress for JT gravity.
Indeed, we have argued in the case of complexity saturation, that the results for JT gravity can be directly translated 
to other higher dimensional black holes. The computation of non-perturbative corrections in JT gravity involves the inclusion of certain fixed energy boundary conditions, which can be reinterpreted as a worldsheet boundary ending on a D-brane \cite{Saad:2019lba,Saad:2019pqd,Blommaert:2019wfy}. 
We can similarly speculate that D-branes reprise their role in the saturation of higher dimensional holographic complexity. The existence of such non-perturbative structures in the black hole spacetime would alter its interior, as expected by the breakdown of bulk reconstruction of interior operators at the saturation of the complexity bound \cite{Akers:2022qdl,Krishnan:2023fnt}.

We note that the saturation of holographic complexity argued for, both in this paper and in the earlier papers on
JT gravity, represents an interesting breakdown of semi-classical gravity through infrared effects. 
Usually, violations of semi-classical gravity are associated with short distance 
physics and strong spacetime curvature, but in this problem the accumulation surface, 
where the bulk of the complexity growth takes place, remains well separated from the black hole singularity 
and the deviation from (semi-)classical physics is instead due to the enormous volume of the extremal surface 
at very late times. Another example of the breakdown of semiclassical gravity computations due to infrared effects is seen in the evaporation of a large near-extremal charged black hole \cite{Brown:2024ajk,Mohan:2024rtn}. In this case, the corrections that invalidate semi-classical calculations at extremely low temperatures come from Schwarzian modes in the emergent JT gravity description of the near-horizon geometry, {\it i.e.} from a region which lies far from the singularity of the black hole.

\bigskip
\leftline{\bf Acknowledgements}
\smallskip
\noindent We are grateful to Chethan Krishnan and Watse Sybesma for useful discussions. Research supported in part by the Icelandic Research Fund under grant 228952-053.  VM is supported by a doctoral grant from The University of Iceland Science Park.  We would also like to thank the Isaac Newton Institute for Mathematical Sciences, Cambridge, for support and hospitality during the programme Black holes: bridges between number theory and holographic quantum information, where work on this project was undertaken. This work was supported by EPSRC grant EP/R014604/1.

\appendix

\section{Examples from asymptotically AdS, dS, and flat spacetime}
\label{flatspacesection}
Our general argument in the main text, that the evaluation of holographic volume complexity at late times reduces to a
calculation of geodesic length in a two-dimensional JT-gravity theory, applies to all static spherically symmetric 
black holes in $d+1$-dimensional Einstein gravity, for any $d\geq 2$. In particular, it does not rely on the presence or
absence of a cosmological constant in the higher-dimensional spacetime, and it also goes through for charged
Reissner-Nordstr\"{o}m black holes. In this appendix we work out specific examples in more detail.

\subsection{(AdS-)Schwarschild Black Hole}
Consider the line element of an eternal AdS black hole in $d+1$-dimensions, 
\bea
\dd s^2=-f(\rho) \dd t^2+\frac{\dd \rho^2}{f(\rho)}+\rho^2 \dd \Omega_{d-1}^2 \,,\qquad 
f(\rho)=1+\frac{\rho^2}{L^2}-\frac{16\pi M \ell_p^{d-1}}{(d-1)\rho^{d-2}}\,,\label{eternalbhmetric}
\eea 
where we have absorbed Newton's constant into the definition of $\ell_p$ as in \eqref{GNabsorbed}.
The horizon of the black hole $\rho_h$ is related to the black hole mass through the relation 
\bea
\frac{16 \pi M \ell_p^{d-1}}{d-1} = \rho_h^{d-2}\Big(\frac{\rho_h^2}{L^2}+1\Big)\, .
\eea
Here $L$ is the $d+1$-dimensional AdS length scale. 
The entropy and the temperature of the black hole are given by
\bea
S=\frac{1}{4} \Big(\frac{\rho_h}{\ell_p}\Big)^{d-1}, \quad 
T=\left.\frac{1}{4 \pi} \frac{\partial f}{\partial \rho}\right|_{\rho=\rho_h}
=\frac{1}{4 \pi \rho_h}\Big(\frac{d}{L^2}\,\rho_h^2+(d-2) \Big) . \label{entropytempeq}
\eea
To obtain the explicit form of the corresponding dilaton potential in \eqref{bulkdilatonactionkineq1} we
start from the Einstein-Hilbert action,
\bea
I=\frac{1}{16 \pi G_{d+1}} \int \dd^{d+1} x \sqrt{-g^{(d+1)}}\Big(R^{(d+1)}+\frac{d(d-1)}{L^2}\Big)\,.
\label{EHaction}
\eea
Performing a spherical reduction in the Weyl frame defined by \eqref{reductionansatzvol1}, we end up with a 
two-dimensional dilaton gravity theory of the form \eqref{bulkdilatonactionkineq1} with
\bea
W(\Phi) =\f{d(d-1)}{L^2}\,\Phi^{-1}+ \f{(d-1)(d-2)}{\ell_p^2}\,\Phi^{\frac{(1+d)}{(1-d)}}\,.\label{dilatonpotential}
\eea
The two-dimensional metric in \eqref{translation}, corresponding to a $d+1$-dimensional AdS-Schwarzschild black hole,
takes a simple form when expressed in terms of the dilaton,
\be\label{adsschwars2dmetriccoompeq}
\xi(r) = \Phi^2+ \f{\ell_p^2}{L^2}\Phi^{\f{2d}{(d-1)}}-\f{16 \pi M\ell_p}{(d-1)}\Phi^{\f{d}{(d-1)}}\,,
\ee
and the dilaton profile is given by
\be
\label{dprofile}
\Phi(r)=\Big(\f{\rho(r)}{\ell_p}\Big)^{(d-1)}=\Big((2d-1)\f{r}{\ell_p}\Big)^\f{(d-1)}{(2d-1)}\,,
\ee
where we have integrated \eqref{varrel} to obtain the relation between $r$ and $\rho$.

Holographic complexity of the higher-dimensional AdS-Schwarzschild black hole reduces to a 
geodesic length in this two-dimensional metric. 
As explained in section \ref{2dsection}, the geodesic computation can be rephrased as a particle scattering off an 
effective potential $V_{\text{eff}}(r) = -\xi(r)$. Plotting the explicit expression, the potential has a local maximum inside
the black hole which corresponds to the radial location of the accumulation surface (see Figure~\ref{effpotentialfig}). 
At the maximum of the potential, the two-dimensional curvature scalar is negative, $R=-\xi''(r_a)=V_{\text{eff}}''(r_a)<0$, 
and it follows that near the accumulation surface the two-dimensional metric reduces to that of a locally AdS$_2$ 
spacetime with an emergent AdS$_2$ length scale that can be obtained from a straightforward analysis of 
\eqref{adsschwars2dmetriccoompeq}. To obtain closed-form expressions, we will work in two limits of the metric:


\noindent \textbf{Large AdS black hole} ($\rho_h \gg L$): In this limit, the accumulation surface sits at 
\bea
r_a = \frac{1}{(2d-1)}\Big(\frac{\rho_a}{\ell_p}\Big)^{2d-1}\ell_p\, , 
\quad \text{with}\quad\rho_a \simeq 2^{-1/d} \rho_h \, .
\eea
Here $\rho_a$ and $\rho_h$ are the radial locations of the accumulation surface and the black hole horizon,
respectively, in the higher dimensional black hole coordinates. 
The late-time rate of growth of the complexity, given by \eqref{laterate}, reduces to
\bea
\frac{\dd C_V}{\dd\tau} \simeq \frac{16\pi M}{(d-1)} \propto ST \,.\label{volgrowthrateeq2}
\eea
Expanding the two-dimensional metric as in section \ref{maincalsec}, 
we obtain an AdS$_2$ black hole with a characteristic length scale $L_2$ and mass parameter $\mu$, with
\bea
L_2 = \sqrt{\frac{2}{\xi^{\prime\prime}(r_a)}} \simeq \frac{1}{d}\Big(\frac{\rho_{a}}{\ell_P}\Big)^{d-1} L 
\qquad \text{and} \qquad \mu \simeq \Big(\frac{\ell_p}{L}\Big)^2\Big(\frac{\rho_{a}}{\ell_p}\Big)^{2d}\,. \label{effadslengthvol}
\eea
Up to a $d$ dependent constant, the emergent AdS$_2$ length scale is given by the Weyl transformation of the higher-dimensional AdS length scale. 

\noindent \textbf{Small AdS black hole} ($\rho_h \ll L$): In this limit, the higher dimensional black hole reduces to an asymptotically flat Schwarzschild black hole and the accumulation surface is located at
\bea
r_a=\frac{1}{(2d-1)}\Big(\frac{d}{2(d-1)}\Big)^{\frac{2d-1}{d-2}} \Big(\frac{\rho_h}{\ell_p}\Big)^{2d-1}\ell_p \, . \label{flatspaceacc}
\eea
As before, the volume complexity is proportional to the length of a geodesic in an emergent two-dimensional theory.
This time around, since the AdS length $L$ is effectively infinite in the small black hole limit,
the characteristic length scale $L_0$ of the higher-dimensional problem is given by the 
Schwarzschild radius $\rho_h$, and equation \eqref{ComplexityLength} becomes
\bea
C_V = \frac{\ell}{\rho_h} \,.\label{complengthflat}
\eea
The late-time growth rate turns out to be
\bea
\frac{\dd C_V}{\dd \tau} \simeq \sqrt{\frac{d-2}{d}}\left(\frac{d}{2(d-1)}\right)^{\frac{(d-1)}{(d-2)}}\frac{\rho_h^{d-2}}{\ell_P^{d-1}} 
\propto ST\,.
\eea
With these conventions, the growth rate of the volume complexity (in Rindler units) is once again proportional 
to the number of degrees of freedom of the black hole.

As in section \ref{maincalsec}, we expand the two-dimensional metric near the accumulation surface. 
This leads to an  AdS$_2$ black hole metric with an emergent length scale given by
\bea
L_2 =\frac{2}{d}\sqrt{\frac{d-1}{d-2}}\Big(\frac{\rho_h}{\ell_p}\Big)^d \ell_p\, , \label{effadslengthvolflat}
\eea
which is again equal to the Weyl transform of the characteristic higher dimensional length scale, up to a $d$ dependent constant.

\subsection{de Sitter-Schwarzschild Black Hole} 
Now consider the $d+1$-dimensional Schwarzschild-de Sitter (SdS) spacetime, whose metric is given by
\bea
\dd s^2=-f(\rho) \dd t^2+\frac{\dd \rho^2}{f(\rho)}+\rho^2 \dd \Omega_{d-1}^2 \,,\qquad f(\rho)=1-\frac{\rho^2}{L^2}-\frac{2 \mu}{\rho^{d-2}}\,.\label{sdsmetric}
\eea 
We choose the mass parameter $\mu$ to lie between 0 and $\mu_N$, where
\be
\mu_N = \frac{L^{d-2}}{d}\left(\frac{d-2}{d}\right)^{\frac{d-2}{2}}\,,
\ee 
corresponds to the Nariai limit. For a mass parameter in this range, 
there is a black hole horizon at $\rho = \rho_h$, and a cosmological horizon at $\rho = \rho_c > \rho_h$ \cite{Balasubramanian:2001nb,Ghezelbash:2001vs}. The SdS metric \eqref{sdsmetric} is a solution of the 
equations of motion of d+1-dimensional Einstein gravity with positive cosmological constant,
\be
I=\frac{1}{16 \pi G_{d+1}} \int \dd^{d+1} x \sqrt{-g^{(d+1)}}\left[R^{(d+1)}-\frac{d(d-1)}{L^2}\right]\,.
\label{EHdsaction}
\ee
A spherical reduction using the ansatz \eqref{reductionansatzvol1}, leads to a two-dimensional theory of the form \eqref{bulkdilatonactionkineq1} with the dilaton potential
\bea
W(\Phi) =-\f{d(d-1)}{L^2}\Phi^{-1}+ \f{(d-1)(d-2)}{\ell_p^2}\Phi^{\frac{1+d}{1-d}}\,.\label{dilatonpotentialds}
\eea
The $d+1$-dimensional de Sitter length provides a reference length scale for the volume complexity,
\bea
C_V = \frac{\ell}{L}\,.
\eea
The geodesic length $\ell$ is calculated using the following two-dimensional metric, 
\be\label{dsschwars2dmetriccoompeq}
\xi = \Phi^2- \f{\ell_p^2}{L^2}\Phi^{\f{2d}{d-1}}-\f{2\mu}{\ell_p^{d-2}}\Phi^{\f{d}{d-1}}\,,\qquad \Phi^{\f{2d-1}{d-1}} = \f{(2d-1)r}{\ell_p}\,,
\ee
where the dilaton profile given by \eqref{dprofile}.
\begin{figure}
  \centering
    \includegraphics[width=0.7\linewidth]{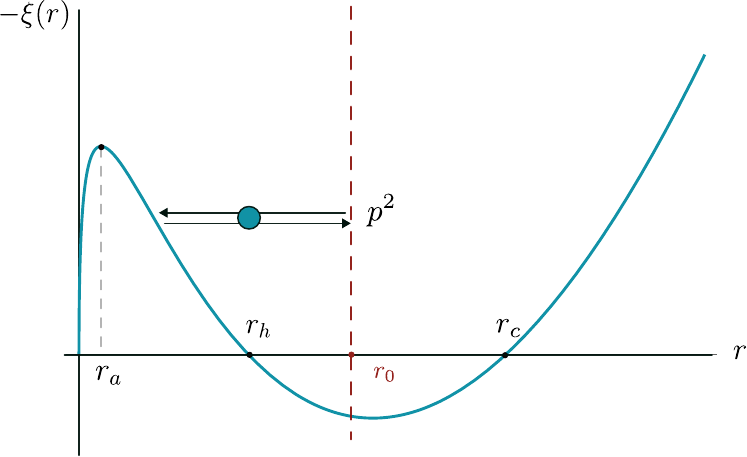}
  \caption{\small Effective potential for spacelike geodesics in SdS spacetime. Here, $r = r_{h,c}$ corresponds to the location of the black hole and cosmological horizons. The geodesic is symmetrically anchored to cutoff surfaces at $r=r_0$ in the static patches on either side of the two-sided black hole.}
  \label{effectivepotentialdsfig}
  \end{figure}
Since our goal is to understand the complexity of the black hole, we will consider extremal volume surfaces 
that pass through the interior of the black hole and are symmetrically anchored at cutoff surfaces in static 
patches on either side.\footnote{Holographic complexity associated with the cosmological horizon was 
studied in \cite{Aguilar-Gutierrez:2024rka}.} 
The effective potential for the corresponding classical particle has a maximum inside the black hole region, as 
shown in Figure~\ref{effectivepotentialdsfig}. The turning point of the geodesic approaches this maximum 
at late times and in the neighbourhood of the critical point one again finds an AdS$_2$ black hole metric with
an emergent AdS$_2$ length scale that descends from the higher-dimensional black hole. 
Since closed-form expressions are unavailable, we do not present the emergent AdS$_2$ length scale here.

\subsection{Reissner-Nordstr\"{o}m Black Hole} 
\begin{figure}
  \centering
    \includegraphics[width=0.7\linewidth]{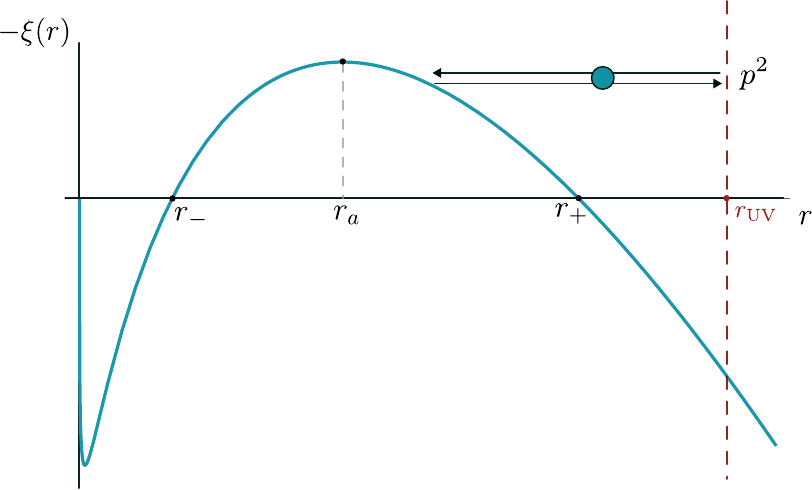}
  \caption{\small Effective potential for spacelike geodesics in Reissner-Nordstr\"{o}m spacetime. 
  Here, $r = r_{\pm}$ corresponds to the two black hole horizons and the geodesics are symmetrically 
  anchored to cutoff surfaces in asymptotic regions on opposite sides of the black hole in a Penrose diagram.}
  \label{effectivepotentialRNfig}
  \end{figure}

The Reissner-Nordstr\"{o}m (RN) black hole in $(d+1)$-dimensions has the metric
\bea
\dd s^2=-f(\rho) \dd t^2+\frac{\dd \rho^2}{f(\rho)}+\rho^2 \dd \Omega_{d-1}^2 \,,
\qquad f(\rho)=1-\frac{16\pi M \ell_p^{d-1}}{(d-1)\rho^{d-2}}+\frac{Q^2\ell_p^{2(d-2)}}{\rho^{2(d-2)}}\,. 
\label{RNmetric}
\eea
We denote the location of the outer and inner black hole horizons by $\rho=\rho_{\pm}$, with $\rho_{+}>\rho_-$.
The metric is a solution to the equations of motion of Einstein-Maxwell theory
\bea
I=\frac{1}{16 \pi G_{d+1}} \int \dd^{d+1} x \sqrt{-g^{(d+1)}}\left[R^{(d+1)}-F_{\mu\nu}F^{\mu\nu}\right]\,,
\label{EHMaction}
\eea
where we have absorbed the coupling constant $1/4g^2$ into the definition of the electromagnetic field strength. 
Spherical reduction using \eqref{reductionansatzvol1} yields a two-dimensional dilaton gravity theory 
coupled to a two-dimensional Maxwell field,
\be
S_\text{bulk}=\frac{1}{16 \pi}\int \dd^2 x \sqrt{-g}\Big(\Phi R^{}-\frac{d}{d-1}\frac{(\nabla\Phi)^2}{\Phi}
+ \f{(d-1)(d-2)}{\ell_p^2}\Phi^{\frac{1+d}{1-d}} -\Phi^3 F^2\Big).
\ee
To bring the action to the form \eqref{bulkdilatonactionkineq1}, we proceed as in \cite{Brown:2018bms} 
and integrate out the electromagnetic field strength. We find that
\be 
\nabla_\alpha \left(\Phi^3 F^{\alpha\beta}\right) = 0 \quad \implies\quad F_{\alpha\beta} =\frac{Q}{\ell_p}  
\sqrt{\frac{(d-1)(d-2)}{2}}\Phi^{-3}\epsilon_{\alpha\beta}\,,
\ee
where $\epsilon_{\alpha \beta}$ is the 2d Levi-Civita tensor. We have chosen the constants in the above 
solution so that the solutions of the resulting 2d theory match the dimensional reduction of \eqref{RNmetric}. 
After introducing an appropriate boundary term, following \cite{Brown:2018bms}, we substitute the on-shell 
expression for $F_{\alpha\beta}$ to obtain \eqref{bulkdilatonactionkineq1} with the following dilaton potential
\be 
W(\Phi) = \f{(d-1)(d-2)}{\ell_p^2}\Phi^{\frac{1+d}{1-d}}-\frac{Q^2(d-1)(d-2)}{\ell_p^2}\Phi^{-3}.
\ee 
Once again, the holographic volume complexity can be expressed in terms of a two-dimensional geodesic length,
\bea
C_V(\tau) = \frac{\ell(\tau)}{\rho_+}\,,
\eea
where we have chosen the radial location of the outer horizon $\rho_+$ as the reference length scale.
The turning points of the geodesics approach an accumulation surface located between the two horizons, 
where the corresponding classical particle effective potential reaches a maximum 
(see Figure~\ref{effectivepotentialRNfig}). As in the previous cases, the late-time complexity growth comes 
from an emergent AdS$_2$ spacetime near the accumulation surface and can be expressed in terms of 
geodesic length in a two-dimensional JT gravity theory. This is true for any Reissner-Nordstr\"{o}m 
black hole. In particular, it does not require the black hole to be near extremal. It is, however, interesting to 
consider the near-extremal limit, on the one hand because it leads to a closed-form expression for 
the late-time complexity growth, and on the other hand because the effective low-energy gravitational 
dynamics in the near-horizon region of near-extremal RN black hole is well known to be governed by a 
two-dimensional JT gravity theory \cite{Navarro-Salas:1999zer,Nayak:2018qej,Brown:2018bms}. 
It is natural to ask whether our JT-gravity prescription for computing volume complexity, when applied 
to a near-extremal RN black hole, differs from the standard near-horizon JT-gravity theory.

To proceed, let us choose $d=3$, and work in the near-extremal limit where
\bea
\rho_+-\rho_-\equiv \delta \rho \ll \rho_+\,.
\eea
The accumulation surface is determined by the critical point of the effective potential,
\bea
r_a \simeq \frac{\rho_+^{5}}{5\ell_p^{4}}-\left(\frac{\rho_+^{4}}{\ell_p^{4}}\right)\frac{\delta \rho}{2}\,. \label{rnacc}
\eea
Expanding the two-dimensional theory near the accumulation surface as in Section~\ref{JTgravitysec}, 
we obtain an AdS$_2$ metric with an emergent length scale,
\bea
L_2 \simeq \frac{\rho_{+}^{3}}{\ell_p^{2}}\label{effadslengthvolRN},
\eea
and the late-time growth rate of complexity is given by
\bea
\frac{\dd C_V}{\dd \tau} \simeq \frac{\delta \rho}{\ell_p^{2}}\, .
\eea
Using
\bea
T = \frac{\delta \rho}{4\pi \rho^2_+} \quad \text{and} \quad S = \frac{\pi \rho^2_+}{G_4}\,,
\eea
we once again find the growth rate to be proportional to the number of degrees of freedom of the black hole
\bea
\frac{\dd C_V}{\dd\tau} \propto ST\, .
\eea
In the near-extremal limit, the location of the accumulation surface in the higher dimensional black hole 
coordinates is $\rho_a\simeq \rho_+ -\frac{\delta \rho}{2}$. This falls inside the near-horizon region governed 
by `standard' JT-gravity theory and, by comparing the Weyl frames and emergent 
length scales of the two JT gravity theories, one finds that they are indeed equivalent in this limit.

\bibliographystyle{JHEP}
\bibliography{refs}

\end{document}